\documentclass[%
 twocolumn,
superscriptaddress,
 amsmath,amssymb,
 aps,
prb,
]{revtex4-2}
\usepackage{graphicx}
\usepackage{dcolumn}
\usepackage{bm}
\usepackage{physics}
\usepackage{soul}
\usepackage{xcolor}

\begin{document}
\setcitestyle{super}
\title{Room temperature optically detected magnetic resonance of single spins in GaN}
\author{Jialun Luo}
 \affiliation{Department of Physics, Cornell University}
\author{Yifei Geng}
\author{Farhan Rana}
\affiliation{School of Electrical and Computer Engineering, Cornell University}
\author{Gregory D. Fuchs}
\email{gdf9@cornell.edu}
\affiliation{
School of Applied and Engineering Physics, Cornell University
}

\date{2023 June 20th}

\maketitle
\textbf{Optically detected magnetic resonance\cite{kohler_magnetic_1993,wrachtrup_optical_1993} (ODMR) is an efficient mechanism to readout the spin of solid-state color centers at room temperature, thus enabling spin-based quantum sensors of magnetic field,\cite{Degen_2008,Taylor_2008,Rondin_2014,gottscholl_spin_2021}  electric field,\cite{dolde_electric-field_2011}  and temperature\cite{acosta_temperature_2010,Toyli_2012,gottscholl_spin_2021} with high sensitivity and broad commercial applicability. The mechanism of room temperature ODMR is based on spin-dependent relaxation between the optically excited states to the ground states, and thus it is an intrinsic property of a defect center.  While the diamond nitrogen-vacancy (NV) center is the most prominent example,\cite{doherty_nitrogen-vacancy_2013,jelezko_single_2006} room temperature ODMR has also been discovered in silicon vacancy centers \cite{widmann_coherent_2015}
and divacancy centers\cite{koehl_room_2011} 
in SiC, and recently in boron vacancy center ensembles\cite{Gottscholl_2020, gao_high-contrast_2021} and unidentified single defects\cite{chejanovsky_single-spin_2021,stern_room-temperature_2022}in hexagonal boron nitride (hBN).
Of these material systems, diamond NV centers are the most technologically important owing to their large (20--30\%) ODMR contrast, long spin coherence, high quantum efficiency, and high brightness.\cite{rondin_magnetometry_2014} Unfortunately, diamond as a substrate is far from being technologically mature. 
For example, diamond is unavailable with high crystalline quality in large-scale wafers and lacks hetero-epitaxial integration with semiconductors for integrated sensor technologies.
Likewise, boron vacancy centers in hBN have large contrast (up to 20\%),\cite{mathur_excited-state_2022, mu_excited-state_2022, gao_high-contrast_2021}  however, they are available only as small flakes, have low quantum efficiency,\cite{reimers_photoluminescence_2020,xu_greatly_2023} and lack a visible zero-phonon line at room temperature.\cite{qian_unveiling_2022} 
Silicon carbide is a technologically mature substrate with recent advances in scalable monolithic integration of color-center-based quantum light sources.\cite{lukin_4h-silicon-carbide--insulator_2020} However, the room temperature ODMR of its defects discovered so far have low contrast (under 1\%).\cite{widmann_coherent_2015, koehl_room_2011, wang_efficient_2017, wang_coherent_2020}
In this work we demonstrate bright single defects in GaN that display large ODMR contrast (up to $\sim$30\%). Because GaN is a mature semiconductor with well-developed electronic technologies already developed, this defect platform is promising for integrated quantum sensing applications.} 

GaN has emerged as a semiconductor of choice for power electronics owing to its wide direct bandgap and high breakdown field.\cite{burk_sic_1999,milligan_sic_2007, chen_gan--si_2017, mishra_algangan_2002} 
Recently, it has also been found to host bright single photon emitters with spectrally narrow photoluminescence (PL) in the visible spectrum.\cite{berhane_bright_2017,berhane_photophysics_2018} These defect centers have zero phonon linewidths of a few meV at room temperature and less than 1 meV at cryogenic temperature.\cite{geng_dephasing_2023,berhane_bright_2017}
These excellent optical properties, combined with the engineerability of GaN make these single-photon emitting defects attractive for on-chip photonics and quantum technologies that require single-photon sources. The atomic structure of these defects has not yet been identified.

In this work, we report that GaN single photon emitters possess spin $S\ge1$ and exhibit high-contrast magnetic field dependent PL and optically detected magnetic resonance (ODMR) at room temperature. Our study reveals at least two distinct groups of defects, each with a distinct ODMR spectrum as well as sign of ODMR contrast. 
This is promising for sensing applications owing to the high ODMR contrast hosted by a mature semiconductor platform, and it is also promising for unraveling the atomic structure of these defect types by providing critical information about the defect orientation within the crystal, spin multiplicity, and sign of the  ODMR response.

\begin{figure*}[!htpb]
    \centering
    \includegraphics[width=0.9\linewidth]{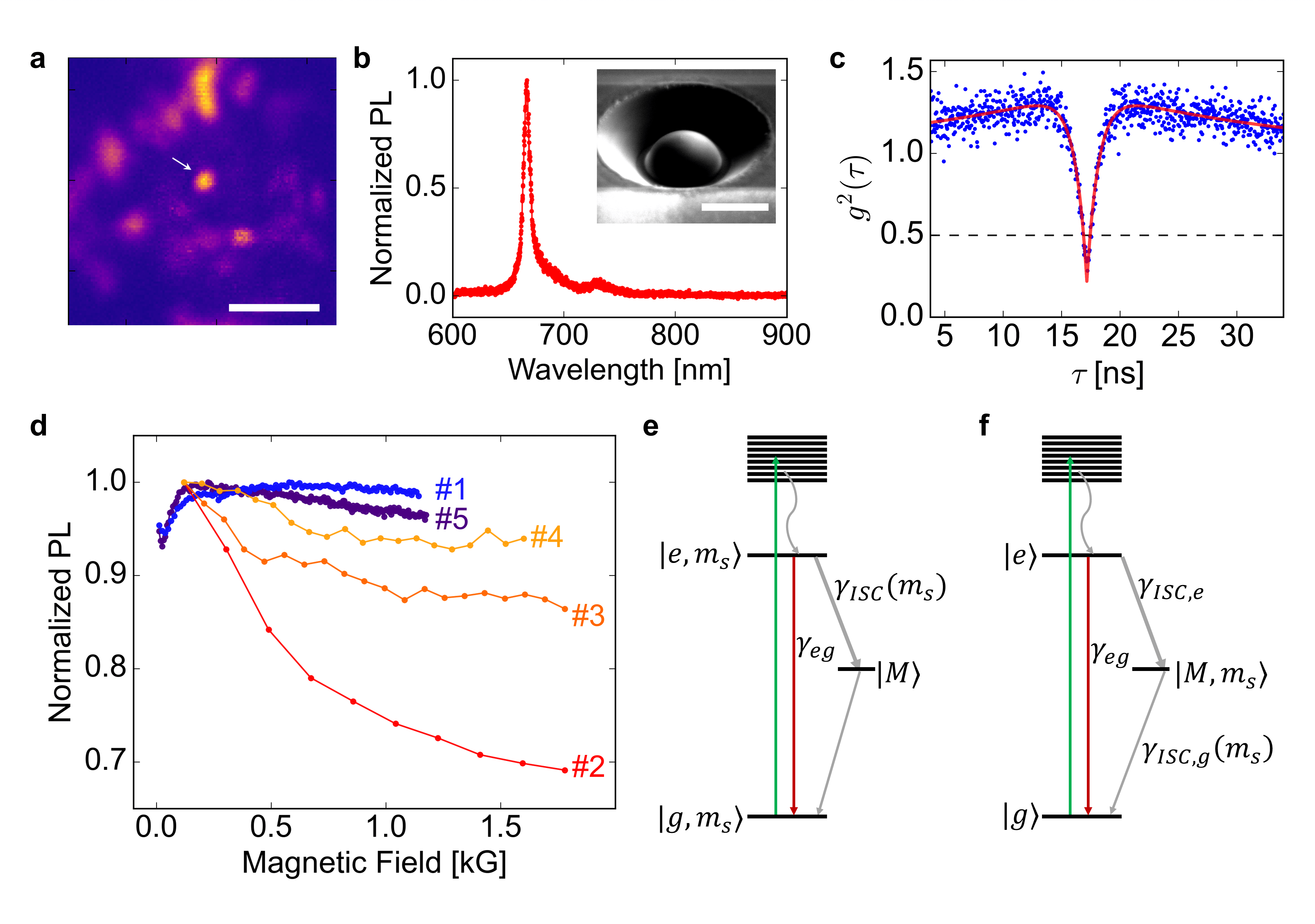}
    \caption{
    \textbf{Optical properties of GaN defects.} 
    (a) Photoluminescence image of an isolated defect (\#2) and its surrounding. The scalebar is 2\,$\mu$m. 
    (b) Optical spectrum of defect \#2. Inset: scanning electron microscope image of a solid-immersion lens carved around the defect. The scalebar is 4\,$\mu$m.
    (c) Second-order photon auto-correlation $g^{(2)}(\tau)$ of defect \#2. The zero-delay autocorrelation $g^{(2)}(0)=0.3<0.5$, which is consistent with a single photon emitter. 
    (d) Magnetic field dependent PL measured with the magnetic field roughly aligned to the $c$-axis of the GaN crystal showing two groups of behavior, as discussed in the text. 
    (e) Minimal level diagram that is consistent with a $S \ge 1$ ground/excited-state spin. The non-radiative intersystem crossing rate $\gamma_\text{ISC}$ into a meta-stable state is spin-dependent.
    (f) Minimal level diagram that is consistent with a $S \ge 1$ metastable state. The non-radiative intersystem crossing rate $\gamma_\text{ISC,g}$ from a meta-stable state and the radiative relaxation rate $\gamma_{eg}$ are spin-dependent.
    }
    \label{fig:SIL_PL_G2}
\end{figure*}

Figure~\ref{fig:SIL_PL_G2}(a-c) detail the typical room temperature optical properties of an  isolated GaN defect used in our study.  The defects are optically separated on the scale of a few micrometers, enabling photon correlation measurements to ensure we examine single defects.  A solid-immersion lens aids in photon collection, with a typical rate of 80~kCounts/s into a 0.9 NA microscope objective when excited with a 532~nm laser with 20~$\mu$W power. Defect \#2 emits most of its PL into a narrow linewidth centered near 667~nm. As noted previously,\cite{berhane_bright_2017,berhane_photophysics_2018,geng_dephasing_2023} not all GaN defects share the same emission energy.  Additionally, while these defects are mainly photo-stable, like most solid-state single photon emitters, these defects suffer some instabilities, including occasional photo-bleaching.  Additional details of the optical properties and photo-stability are discussed in the supplementary information. 

A simple method of screening a particular defect for spin-dependent optical properties is measuring its magnetic field dependent PL (magneto-PL).\cite{epstein_anisotropic_2005,exarhos_magnetic-field-dependent_2019}  Although the specific magneto-PL response depends on the angle of the magnetic field with respect to the defect spin quantization axis, we select the GaN $c$-axis as a potential direction of high symmetry.  The result for five individual defects is shown in Fig.~\ref{fig:SIL_PL_G2}(d). We immediately notice that the defects fall into two groups of behavior.  In the first group, there is a $\sim$7\%  dip in PL at low magnetic fields, followed by an increase of PL to saturation (\#1 and \#5, which we label group I).  In the second group, the PL falls monotonically with magnetic field, showing up to a 30\% change in PL (\#2 -- \#4, which we label group II).   

We proceed under the initial assumption that these GaN defect groups have a spin-dependent PL mechanism similar to that of the diamond nitrogen-vacancy center, in which a spin-state-dependent intersystem crossing can occur from the excited states to a metastable state (Fig.~\ref{fig:SIL_PL_G2}(e)), which ultimately creates spin-dependent PL contrast. \cite{epstein_anisotropic_2005}  It is also possible to obtain spin-dependent PL even if the ground and excited states are spin singlets or doublets if there is spin-dependent relaxation from a $S \ge 1$ metastable states (Fig.~\ref{fig:SIL_PL_G2}(f)).\cite{lee_readout_2013,exarhos_magnetic-field-dependent_2019}  In both cases, the precise spin contrast results from a competition between radiative and non-radiative relaxation rates, branching ratios, and optical pumping rates. While the full characterization of the GaN defect optical cycle is beyond the scope of this work, we re-examine some details of the spin-dependent optical cycle below.  

Regardless of the specific mechanism of spin-dependent PL contrast, magneto-PL originates from Zeeman-induced spin state mixing between spin states with different average PL rates.
This mechanism is relevant to systems with electronic spin $S\ge1$.
The ground state Hamiltonian of a spin system with $S\ge1$ in a magnetic field $B$ is given by
\begin{equation}
\mathcal{H} = D \qty(S_z^2 - \frac{1}{3} S(S+1)) + E (S_x^2-S_y^2) + g \mu_B \Vec{S}\cdot \Vec{B},
\label{eq:hamiltonian}
\end{equation}
where $S$ is the electronic spin operator, $g$ the electronic $g$-factor, $\mu_B$ the Bohr magneton, $D$ and $E$ together the zero-field interaction parameters.
An angle between the external field $B$ and the spin quantization axis introduces off-diagonal large matrix elements between the spin eigenstates, mixing them.
The spin eigenstates can also mix at low fields if $E \ne 0$.

Returning to the magneto-PL measurements of Fig.~\ref{fig:SIL_PL_G2}(d), the group-I magneto-PL response suggests a Zeeman-induced spin degeneracy at low magnetic fields, suggesting $S\ge 1$ with a value of $D$ of only a few hundred megahertz, depending somewhat on the direction of the magnetic field.  Additionally, it suggests that optical pumping puts defects in this group into a state with higher PL, while spin mixing reduces the overall PL; a situation similar to diamond NV centers.  While group II defects also must have $S\ge 1$, in contrast to group I defects, the magneto-PL is monotonically decreasing. This could be explained by a very large misalignment angle of the magnetic field with respect to the defect symmetry axis, or by an opposite contrast of PL, where the optically polarized state has lower PL than the spin-mixed state. 

\begin{figure}[hb]
\centering
\includegraphics[width=1 \linewidth]{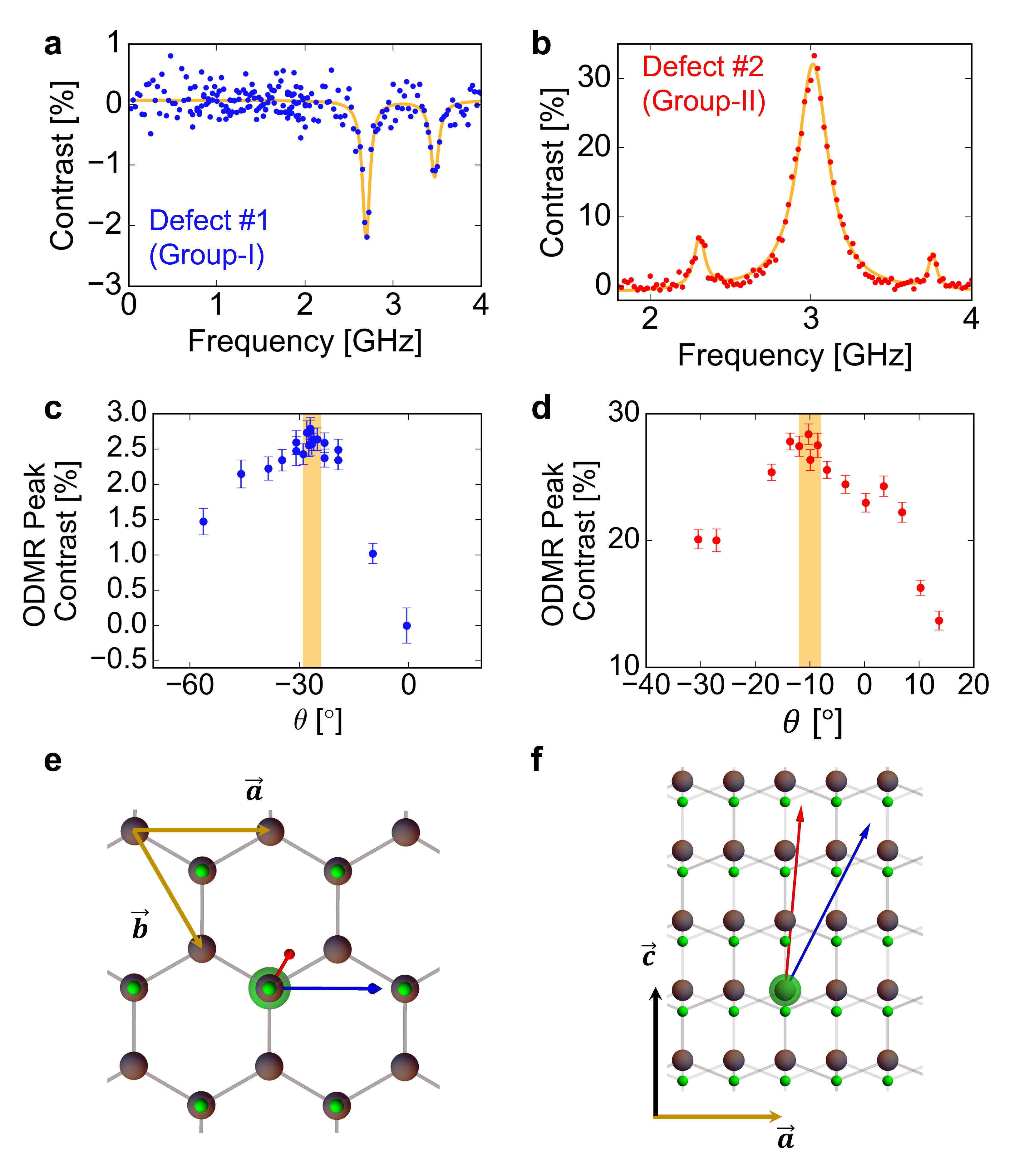}
\caption{\textbf{Optically detected magnetic resonance.} (a) and (b) show ODMR signals of defects \#1 and \#2 at $B=1$~kG respectively. (c) and (d) show the dependence of the ODMR peak contrasts of defects \#1 and \#2 on the alignment between the magnetic field and the crystal $c$-axis. (e) and (f) visualize the spin quantization axes with respect to the lattice. The blue arrow represents defect \#1 and the red represents defect \#2.}
\label{fig:ODMRSampleTraces}
\end{figure}

Having confirmed that both groups of individual defects have spins with $S\ge1$ and a spin-dependent optical cycle, we study the spin-resonant transitions and spin Hamiltonian by measuring continuous wave (cw-) ODMR.
To study the spin resonance, we continuously drive a microwave magnetic field, optically pump the defect optical transition, and count the emitted PL.  
Figure~\ref{fig:ODMRSampleTraces} shows the resulting cw-ODMR traces at $B=1$~kG for a group-I (\#1) and a group-II (\#2) defect.
We immediately notice that the two groups have an opposite sign of ODMR contrast, as suggested by the magneto-PL, with group-I defects showing negative cw-ODMR contrast and group-II defects showing positive cw-ODMR contrast. 
We also notice that the group-I defect has a modest contrast of $\sim$2\% at this driving power, while the group-II defect has a $\sim$30\% contrast for one of the three resonance features, with smaller contrast for two other features. 
The resonances are each well-fit by Lorentzian lineshapes.

A key input for establishing the identity of a new defect is its spin quantization axis.  Having discovered a reliable cw-ODMR signal on multiple GaN single defects, we now make the assumption that the cw-ODMR contrast will be largest when we align the external magnetic field along the $z$-axis defined by Eqn.~\ref{eq:hamiltonian}. 
A misaligned static field will mix the spin eigenstates, which will reduce the cw-ODMR contrast if the fluorescence contrast mechanism is tied to $|m_s|$ as it is for the diamond NV center.  To test this we systematically vary the polar angle $\theta$ with respect to the $c$-axis of the crystal, and then the azimuthal angle $\phi$, which is measured with respect to the $a$-lattice vector of GaN. 
Fig.~\ref{fig:ODMRSampleTraces}(c) and (d) show the ODMR contrast for defect \#1 and \#2, respectively, as a function of $\theta$, while the corresponding data as a function of $\phi$ is shown in Fig.~\ref{fig:extFig:defect1and2AngleData} in the supplementary materials. We find that 
the spin quantization axis for the group-I defect \#1 forms a  $\sim$27-degree angle with the GaN crystal $c$-axis, with an in-plane component points along the $a$-axis.
For the group-II defect \#2 we find a spin quantization axis approximately 10 degrees away from the $c$-axis, and an in-plane component along the $a$-axis. 
Neither spin quantization axis matches a vector between a lattice site and its nearest few neighbors, suggesting the involvement of interstitial atoms (Fig.~\ref{fig:ODMRSampleTraces}(e)-(f)).

\begin{figure}[!htpb]
\centering
\includegraphics[width=\linewidth]{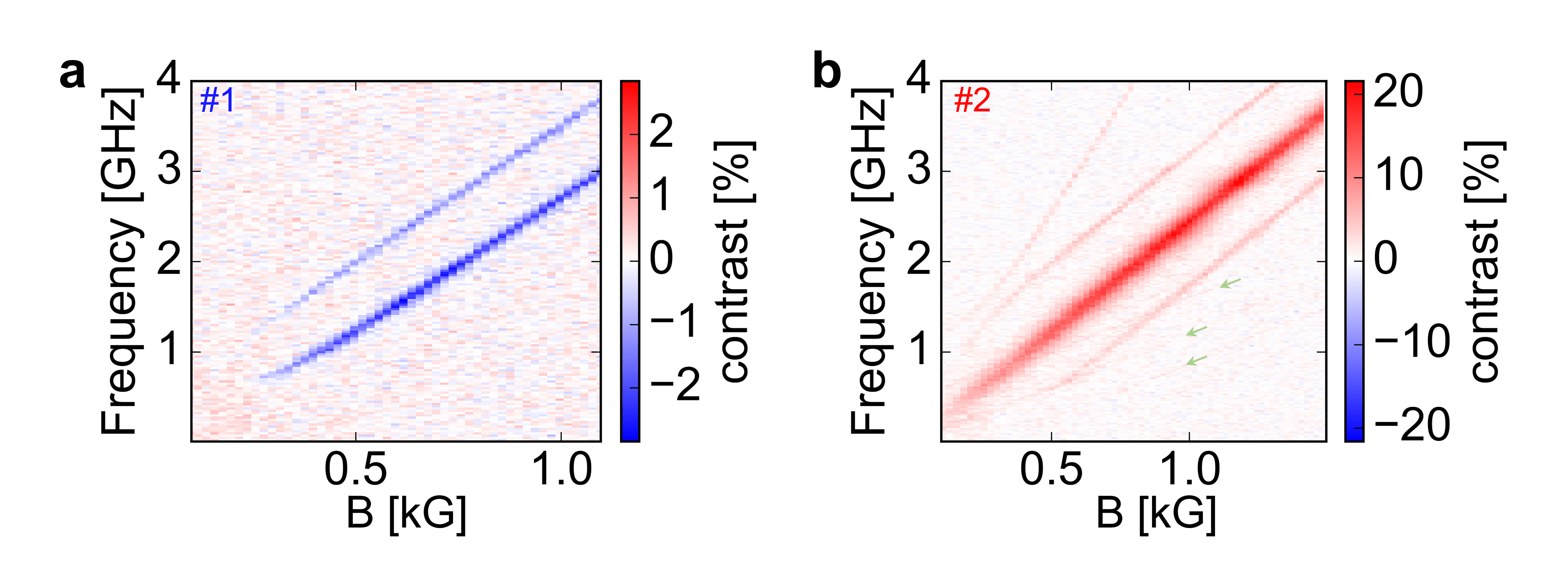}
\caption{\textbf{cw-ODMR spectrum as a function of magnetic field.} The magnetic-field-dependent ODMR signals for defects (a) \#1 and (b) \#2. Two spin resonances can be seen on the group-I defect \#1 and four can be seen on the group-II defect \#2. Note that the three faint lines dispersing with $g=1$ indicated by the arrows are harmonic replica artifacts due to the microwave power amplifier nonlinearity.
}
\label{fig:ODMRvsB}
\end{figure}

Now we study the Zeeman effect on the spin levels.  First, we align our set-up so that $\vec{B}$ is parallel to the direction of the largest ODMR contrast discussed above and record ODMR as a function of $B$.  Under these conditions we assume $B = B_Z$ from Eqn.~\ref{eq:hamiltonian}.
Figure \ref{fig:ODMRvsB} shows the resulting cw-ODMR data from defect \#1 (group I) and defect \#2 (group II) from $100$\,G to $1500$\,G.
The most visible spin resonances disperse with a $g$-factor $g=2$, confirming that we study electronic spins.

First focusing on defect \#1, we see two transitions of unequal contrast that appear at $B \gtrsim 250$\,G.
The lack of cw-ODMR contrast at low magnetic fields suggests a mixing between the spin eigenstates that leads to the suppression of spin contrast. 
If we assume a minimum spin multiplicity to explain the two transitions, $S=1$, then this data can be described by Eqn.~\ref{eq:hamiltonian} with $D\approx E\approx 389$~MHz.  An overlay of the fitted spin transitions is shown in the supplementary materials (Fig.\ref{fig:extFig:allODMRSpectra}(a)).  
Under these conditions at low magnetic fields, the zero-field spin eigenstates would indeed be strongly mixed, thus suppressing spin-dependent optical contrast.  We note, however, that this scenario does not explain why the two transitions have unequal contrast, which may relate to dynamics of the optical cycle that have not been revealed by these measurements.  Additionally,  we find that the model deviates from the data at the lowest magnetic fields, which may point to other physics not contained in a toy model of a single electronic spin-1.  For example, the Ga and N atoms that surround the defect all have a nonzero nuclear spin, which may interact very strongly with this defect and thus potentially explain a deviation from a simple electronic model.  Additionally, we note that group-I defects are rare compared to those in group II.  While we observed magneto-PL for two defects in this group, one of those stopped being optically active, and thus defect \#1 is the only group-I defect that we have been able to record ODMR.  More information can be found in the supplementary information.

Next we examine the field-dependent cw-ODMR of defect \#2, which has the same cw-ODMR spectrum as all of the group-II defects that we studied. Data for other defects can be found in Fig.~\ref{fig:extFig:allODMRSpectra} in the supplementary information. This defect shows three spin transitions that disperse with $g=2$, making spin $S = 3/2$ a minimal model assuming that there is an ODMR contrast mechanism for all $\Delta m_s = 1$ transitions. Again we note that the three transitions have unequal contrast. The strongest cw-ODMR feature extrapolates to zero frequency at zero field within experimental uncertainty, suggesting that it is due to a transition between $\ket{m_s = -\frac{1}{2}}$ and $\ket{m_s = +\frac{1}{2}}$ in this picture.
In addition to the $g=2$ resonance, we also see a 4th resonance that disperses with $g=4$.  Additionally, at $B\sim 300$\,G 
and  $f_\text{mw}=1.5$~GHz, this feature appears to have an avoided-crossing with the highest frequency $g=2$ spin resonance.
Although a $g=4$ resonance can be explained by a $\Delta m_s = 2$ spin transition, that scenario does not give rise to an avoided-crossing, suggesting that a toy electronic model based on Eqn.~\ref{eq:hamiltonian} is insufficient to describe this spin system if the magnetic field is aligned along the symmetry axis.  If we ignore the $g = 4$ resonant line, these transitions are well-described for $B>0.5$ kG by a $S=3/2$ model with $D =$ 368 MHz and $E =$ 0.

\begin{figure}[!hpbt]
\centering
\includegraphics[width=\linewidth]{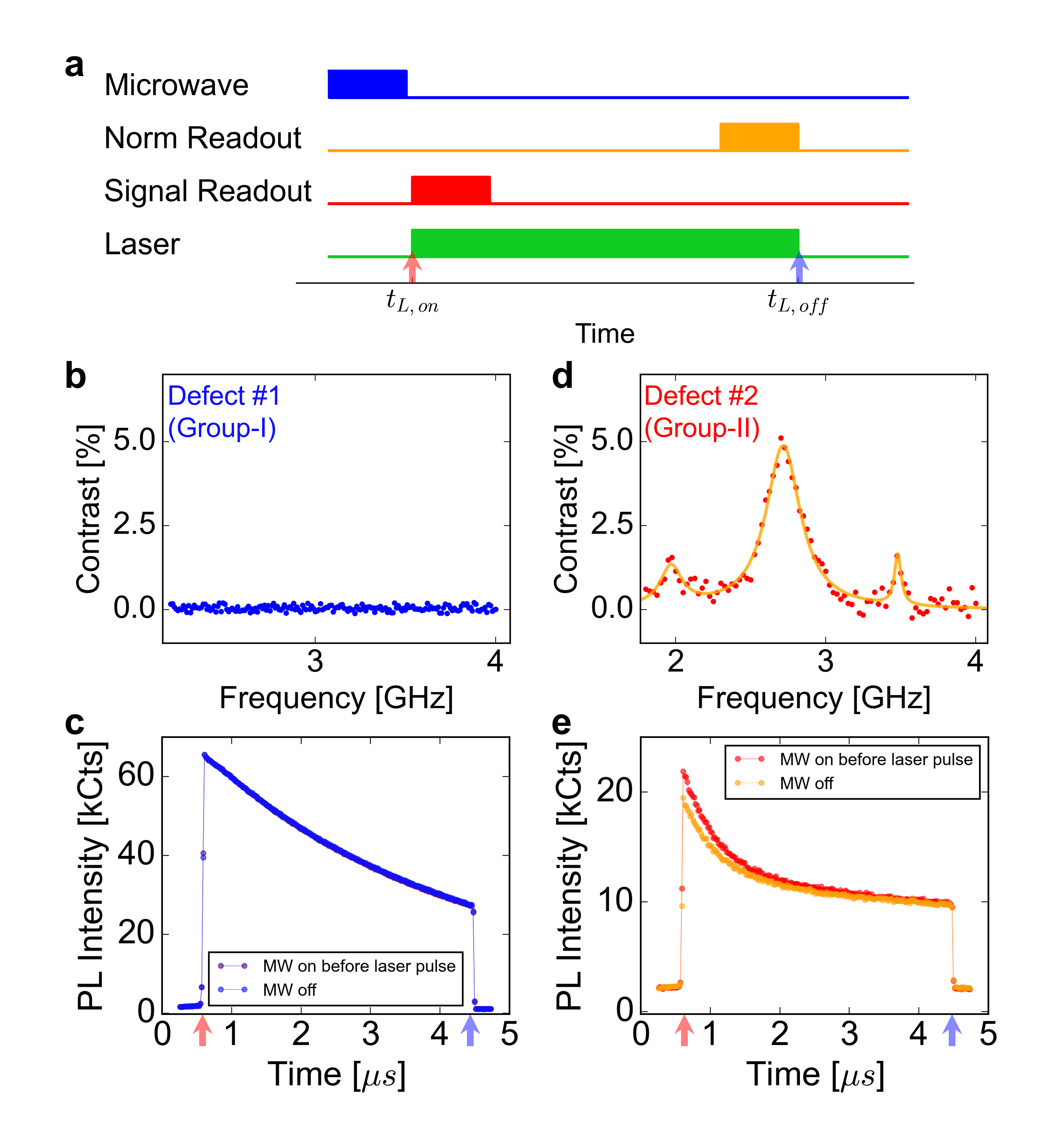}
\caption{\textbf{Spin-dependent optical dynamics} (a) Timing diagram of a single measurement cycle of the pulsed ODMR and time-resolved PL schemes. The microwave pulse (blue) is applied before turning on the laser (green). 
After the optical readout, the laser is turned off to relax all population out of the metastable state before the microwave pulse turns back on in the next cycle.
The optical detection is via either integrated counting during the signal and normalization windows (red and orange, respectively), or time-resolved single photon counting. 
(b) Pulsed ODMR measurement of defect \#1. (c) Time-resolved PL of defect \#1, with and without a microwave pulse applied prior to the laser pulse. Note the two data sets sit right on top of each other. (d) Pulsed ODMR measurement of defect \# 2. (e) Time-resolved PL of defect \#2, with and without a microwave pulse applied prior to the laser pulse. The red and blue arrows in (a, c, e) represent the times when the laser turns on and off, respectively. 
}
\label{fig:TimeResolvedPL}
\end{figure}
Finally, we return to the question of whether the spins associated with these defects are in the ground-state and excited-state manifold as in the case of the diamond NV center (Fig.~\ref{fig:SIL_PL_G2}(e)), or whether they are associated with a metastable state (Fig.~\ref{fig:SIL_PL_G2}(f)) as in the case of the diamond ST1 defect.\cite{lee_readout_2013}
To clarify that assignment, we perform both pulsed ODMR and time-resolved single photon counting experiments with separate microwave spin manipulation and optical excitation. 
The pulse timings are detailed in the supplementary information.
If we manipulate a ground-state spin, then we expect the pulsed ODMR scheme shown in Fig.~\ref{fig:TimeResolvedPL}(a) to result in a visible spin resonance.  However, if there is no contrast, then we can assign the cw-ODMR response to a metastable spin state.

We start with pulsed ODMR of defect \#1 from group I, shown in Fig.~\ref{fig:TimeResolvedPL}(b).  We observe no spin resonance response, with the noise floor of our integration at the level of 0.2\%.  Comparing this figure to the $\sim$2\% contrast that we observed for cw-ODMR, we conclude that this defect likely has a ground-state/excited-state singlet or a ground-state doublet with no ODMR contrast.  Thus, the $S\ge 1$ spin state that gives rise to cw-ODMR must reside in a metastable state.  We confirm our conclusion that ground-state microwave preparation has no impact on the PL using direct time-resolved single photon counting (Fig.~\ref{fig:TimeResolvedPL}(c)).  We see that after turning on the laser, defect \#1 has a microsecond-timescale reduction in PL as a function of time, however, we note no difference between the curves generated by a laser pulse alone and a microwave pulse followed by a laser pulse.  These data, along with measurements of $g^{(2)}$ of this defect that shows photon bunching (see the supplementary information), support the existence of a metastable state, and are consistent with the picture of a $S\ge 1$ metastable state.  Further work will be necessary to pin down all the rates in the optical cycle; however, these measurements all point to an optical cycle like that schematically shown in Fig.~\ref{fig:SIL_PL_G2}(f).

Next we repeat this series of measurements on a member of group II, defect \#2, and find the opposite result in Fig.~\ref{fig:TimeResolvedPL}(d).  Here we find visible pulsed ODMR contrast, confirming that the cw-ODMR measurements are the result of a ground-state spin.  Interestingly, while the pulsed ODMR contrast is lower than the cw-ODMR owing to different details of the measurement protocol, we see that the same ratio of contrast between the the three $\Delta m_s = 1$ transitions is preserved.  This suggests that there are non-trivial spin-dependent intersystem crossing rates, and in particular they are not proportional to $\abs{m_s}$ as in the case of diamond NV centers.  We perform the time-resolved PL measurement of defect \#2 as before, with the microwave pulse tuned to the largest-contrast resonance.  As expected, we see a noticeably larger initial PL response when we manipulate the ground-state spin before the laser pulse than when we do not, with a contrast lasting for $\sim$2 $\mu$s.  While this experiment does not establish all the details of the optical cycle and spin, it is consistent with a level diagram and dynamics as shown in Fig.~\ref{fig:SIL_PL_G2}(e).

In conclusion, we report high-contrast optically detected spin resonance of GaN single defect spins at room temperature. 
We find two distinct defect groups that we categorize based on their magnto-PL and ODMR spectra. They display complex optical cycles and spin resonance behavior that will require further investigation to understand fully, however, this work establishes key facts of these defect groups.  The first group has a small negative ODMR contrast, with spin at least $S = 1$ in its metastable state to explain the experimental results.  The second group has a large (up to $30\%$) positive ODMR contrast, with a complicated ground-state spin Hamiltonian including at least $S = 3/2$.  Additionally, through angle-dependent cw-ODMR measurements, we establish a spin quantization axis in terms of the magnetic field angle with largest ODMR contrast.  
The spin quantization axes of both groups do not connect neighboring GaN lattice sites, suggesting the involvement of interstitials.  Beyond providing critical new clues to help identify these high performance single photon emitters, our findings are promising as the basis for magnetic sensing technologies using defect fluorescence based a  mature optoelectronic semiconductor platform.  

\section*{Methods}
\textbf{Sample preparation.}
We study a GaN sample commercially available from the Xiamen Powerway Advanced Material Co., Limited, China. A 4~$\mu$m-thick layer of GaN is grown on a 430~$\mu$m-thick sapphire wafer by hydride vapor phase epitaxy (HVPE). The GaN is Fe-doped to make it semi-insulating. 
We pre-select GaN defects using our home-built scanning laser confocal microscope.
We check that the PL spectra of the defects are consistent with the ones previously reported\cite{berhane_bright_2017,berhane_photophysics_2018,geng_dephasing_2023} and verify that they are single photon emitters by measuring the photon auto-correlation $g^{(2)}$.
GaN is a high-index material with $n\sim2.4$, which leads to a low fraction of PL leaving the material.  To enhance photon collection,  we use focused-ion-beam milling to carve out a 4~$\mu$m-diameter hemisphere-shaped solid-immersion lens (SIL) on the pre-selected defects.
We conduct all measurements at room temperature.

\textbf{Magneto-PL.}
We use a 50.4\,mm-diameter 50.4\,mm-long cylindrical neodymium iron boron permanent magnet to apply magnetic fields to the sample.  To adjust the magnetic field amplitude and direction, we move the magnet on a motorized translation stage, having calibrated the magnetic field against magnet position. The details of the magnet setup are described in the supplementary materials.

\textbf{Continuous-wave ODMR (cw-ODMR).}
To drive spin resonance, a copper microwire is lithographically patterned near the SILs containing the defects of interest.
The details of the microwave set-up are described in the supplementary materials.
We drive about 20\,dBm of microwave power to induce the spin resonances and excite the defects with an optical power of 15--20~$\mu$W.

\textbf{Pulsed measurements.}
Figure~\ref{fig:TimeResolvedPL}(a) shows the pulse scheme in a measurement cycle for pulsed ODMR and time-resolved PL measurements. 
The details of the timing can be found in the supplementary information Fig.~\ref{fig:extFig:pulsedTiming}.
In both schemes, we apply microwaves before we excite the defects for optical readouts, and
we turn off the laser for a sufficient time before we apply microwaves again in the next cycle to allow relaxation from all populations to the ground state.

Supplementary Figure~\ref{fig:extFig:pulsedTiming}(a) shows the timings of a cycle of pulsed ODMR measurement. After the optical pulse has been off for 3~$\mu$s, the microwave pulse in the next cycle is turned on for 2~$\mu$s and off 65~ns before the laser excitation. We read the PL for 2~$\mu$s  after the microwave turns off and normalize it to the PL registered after the laser has repolarized the system for 8~$\mu$s.  This sequence is designed to distinguish between a ground-state and a meta-stable state spin.

To measure time-resolved PL, we apply a microwave pulse tuned to the largest contrast resonance frequency for 1~$\mu$s before the laser turns on as depicted in Fig.~\ref{fig:extFig:pulsedTiming}(b), and we allow the system to relax for 1.5~$\mu$s before applying a microwave pulse again in the next cycle.
The optical detection is done by a time-correlated single-photon-counting (TCSPC) module that is triggered by a synchronization pulse when the laser turns on at $t_{L,on}$ in each pulse cycle.
This way, we record the photon arrival times relative to the laser excitation time. The histogram of photon arrival times gives the time-resolved PL.

\section*{Acknowledgements}
We thank Len van Deurzen, Debdeep Jena, and Huili Grace Xing for useful discussions and for supplying the GaN substrates. We thank Brendan McCullian, Nikhil Mathur, Anthony D'Addario, and Johnathon Kuan for very helpful discussions on the physics and microwave experiments. This work was supported by the Cornell Center for Materials Research (CCMR), an NSF Materials Research Science and Engineering Center (DMR-1719875).  Preliminary work was supported by the NSF TAQS program (ECCS-1839196).  We also acknowledge support through the Cornell Engineering Sprout program.  This work was performed in part at the Cornell NanoScale Science \& Technology Facility (CNF), a member of the National Nanotechnology Coordinated Infrastructure (NNCI), which is supported by the NSF (Grant NNCI-1542081)

\bibliography{JialunODMRPaperCitations}

\begin{thebibliography}{36}%
\makeatletter
\providecommand \@ifxundefined [1]{%
 \@ifx{#1\undefined}
}%
\providecommand \@ifnum [1]{%
 \ifnum #1\expandafter \@firstoftwo
 \else \expandafter \@secondoftwo
 \fi
}%
\providecommand \@ifx [1]{%
 \ifx #1\expandafter \@firstoftwo
 \else \expandafter \@secondoftwo
 \fi
}%
\providecommand \natexlab [1]{#1}%
\providecommand \enquote  [1]{``#1''}%
\providecommand \bibnamefont  [1]{#1}%
\providecommand \bibfnamefont [1]{#1}%
\providecommand \citenamefont [1]{#1}%
\providecommand \href@noop [0]{\@secondoftwo}%
\providecommand \href [0]{\begingroup \@sanitize@url \@href}%
\providecommand \@href[1]{\@@startlink{#1}\@@href}%
\providecommand \@@href[1]{\endgroup#1\@@endlink}%
\providecommand \@sanitize@url [0]{\catcode `\\12\catcode `\$12\catcode
  `\&12\catcode `\#12\catcode `\^12\catcode `\_12\catcode `\%12\relax}%
\providecommand \@@startlink[1]{}%
\providecommand \@@endlink[0]{}%
\providecommand \url  [0]{\begingroup\@sanitize@url \@url }%
\providecommand \@url [1]{\endgroup\@href {#1}{\urlprefix }}%
\providecommand \urlprefix  [0]{URL }%
\providecommand \Eprint [0]{\href }%
\providecommand \doibase [0]{https://doi.org/}%
\providecommand \selectlanguage [0]{\@gobble}%
\providecommand \bibinfo  [0]{\@secondoftwo}%
\providecommand \bibfield  [0]{\@secondoftwo}%
\providecommand \translation [1]{[#1]}%
\providecommand \BibitemOpen [0]{}%
\providecommand \bibitemStop [0]{}%
\providecommand \bibitemNoStop [0]{.\EOS\space}%
\providecommand \EOS [0]{\spacefactor3000\relax}%
\providecommand \BibitemShut  [1]{\csname bibitem#1\endcsname}%
\let\auto@bib@innerbib\@empty
\bibitem [{\citenamefont {Köhler}\ \emph {et~al.}(1993)\citenamefont
  {Köhler}, \citenamefont {Disselhorst}, \citenamefont {Donckers},
  \citenamefont {Groenen}, \citenamefont {Schmidt},\ and\ \citenamefont
  {Moerner}}]{kohler_magnetic_1993}%
  \BibitemOpen
  \bibfield  {author} {\bibinfo {author} {\bibfnamefont {J.}~\bibnamefont
  {Köhler}}, \bibinfo {author} {\bibfnamefont {J.~a. J.~M.}\ \bibnamefont
  {Disselhorst}}, \bibinfo {author} {\bibfnamefont {M.~C. J.~M.}\ \bibnamefont
  {Donckers}}, \bibinfo {author} {\bibfnamefont {E.~J.~J.}\ \bibnamefont
  {Groenen}}, \bibinfo {author} {\bibfnamefont {J.}~\bibnamefont {Schmidt}},\
  and\ \bibinfo {author} {\bibfnamefont {W.~E.}\ \bibnamefont {Moerner}},\
  }\bibfield  {title} {\bibinfo {title} {Magnetic resonance of a single
  molecular spin},\ }\href {https://doi.org/10.1038/363242a0} {\bibfield
  {journal} {\bibinfo  {journal} {Nature}\ }\textbf {\bibinfo {volume} {363}},\
  \bibinfo {pages} {242} (\bibinfo {year} {1993})}\BibitemShut {NoStop}%
\bibitem [{\citenamefont {Wrachtrup}\ \emph {et~al.}(1993)\citenamefont
  {Wrachtrup}, \citenamefont {von Borczyskowski}, \citenamefont {Bernard},
  \citenamefont {Orrit},\ and\ \citenamefont {Brown}}]{wrachtrup_optical_1993}%
  \BibitemOpen
  \bibfield  {author} {\bibinfo {author} {\bibfnamefont {J.}~\bibnamefont
  {Wrachtrup}}, \bibinfo {author} {\bibfnamefont {C.}~\bibnamefont {von
  Borczyskowski}}, \bibinfo {author} {\bibfnamefont {J.}~\bibnamefont
  {Bernard}}, \bibinfo {author} {\bibfnamefont {M.}~\bibnamefont {Orrit}},\
  and\ \bibinfo {author} {\bibfnamefont {R.}~\bibnamefont {Brown}},\ }\bibfield
   {title} {\bibinfo {title} {Optical detection of magnetic resonance in a
  single molecule},\ }\href {https://doi.org/10.1038/363244a0} {\bibfield
  {journal} {\bibinfo  {journal} {Nature}\ }\textbf {\bibinfo {volume} {363}},\
  \bibinfo {pages} {244} (\bibinfo {year} {1993})}\BibitemShut {NoStop}%
\bibitem [{\citenamefont {Degen}(2008)}]{Degen_2008}%
  \BibitemOpen
  \bibfield  {author} {\bibinfo {author} {\bibfnamefont {C.~L.}\ \bibnamefont
  {Degen}},\ }\bibfield  {title} {\bibinfo {title} {Scanning magnetic field
  microscope with a diamond single-spin sensor},\ }\href
  {https://doi.org/10.1063/1.2943282} {\bibfield  {journal} {\bibinfo
  {journal} {Applied Physics Letters}\ }\textbf {\bibinfo {volume} {92}},\
  \bibinfo {pages} {243111} (\bibinfo {year} {2008})}\BibitemShut {NoStop}%
\bibitem [{\citenamefont {Taylor}\ \emph {et~al.}(2008)\citenamefont {Taylor},
  \citenamefont {Cappellaro}, \citenamefont {Childress}, \citenamefont {Jiang},
  \citenamefont {Budker}, \citenamefont {Hemmer}, \citenamefont {Yacoby},
  \citenamefont {Walsworth},\ and\ \citenamefont {Lukin}}]{Taylor_2008}%
  \BibitemOpen
  \bibfield  {author} {\bibinfo {author} {\bibfnamefont {J.~M.}\ \bibnamefont
  {Taylor}}, \bibinfo {author} {\bibfnamefont {P.}~\bibnamefont {Cappellaro}},
  \bibinfo {author} {\bibfnamefont {L.}~\bibnamefont {Childress}}, \bibinfo
  {author} {\bibfnamefont {L.}~\bibnamefont {Jiang}}, \bibinfo {author}
  {\bibfnamefont {D.}~\bibnamefont {Budker}}, \bibinfo {author} {\bibfnamefont
  {P.~R.}\ \bibnamefont {Hemmer}}, \bibinfo {author} {\bibfnamefont
  {A.}~\bibnamefont {Yacoby}}, \bibinfo {author} {\bibfnamefont
  {R.}~\bibnamefont {Walsworth}},\ and\ \bibinfo {author} {\bibfnamefont
  {M.~D.}\ \bibnamefont {Lukin}},\ }\bibfield  {title} {\bibinfo {title}
  {High-sensitivity diamond magnetometer with nanoscale resolution},\ }\href
  {https://doi.org/10.1038/nphys1075} {\bibfield  {journal} {\bibinfo
  {journal} {Nature Physics}\ }\textbf {\bibinfo {volume} {4}},\ \bibinfo
  {pages} {810} (\bibinfo {year} {2008})}\BibitemShut {NoStop}%
\bibitem [{\citenamefont {Rondin}\ \emph
  {et~al.}(2014{\natexlab{a}})\citenamefont {Rondin}, \citenamefont {Tetienne},
  \citenamefont {Hingant}, \citenamefont {Roch}, \citenamefont {Maletinsky},\
  and\ \citenamefont {Jacques}}]{Rondin_2014}%
  \BibitemOpen
  \bibfield  {author} {\bibinfo {author} {\bibfnamefont {L.}~\bibnamefont
  {Rondin}}, \bibinfo {author} {\bibfnamefont {J.~P.~P.}\ \bibnamefont
  {Tetienne}}, \bibinfo {author} {\bibfnamefont {T.}~\bibnamefont {Hingant}},
  \bibinfo {author} {\bibfnamefont {J.~F.~F.}\ \bibnamefont {Roch}}, \bibinfo
  {author} {\bibfnamefont {P.}~\bibnamefont {Maletinsky}},\ and\ \bibinfo
  {author} {\bibfnamefont {V.}~\bibnamefont {Jacques}},\ }\bibfield  {title}
  {\bibinfo {title} {Magnetometry with nitrogen-vacancy defects in diamond},\
  }\href {https://doi.org/10.1088/0034-4885/77/5/056503} {\bibfield  {journal}
  {\bibinfo  {journal} {Rep Prog Phys}\ }\textbf {\bibinfo {volume} {77}},\
  \bibinfo {pages} {56503} (\bibinfo {year} {2014}{\natexlab{a}})}\BibitemShut
  {NoStop}%
\bibitem [{\citenamefont {Gottscholl}\ \emph {et~al.}(2021)\citenamefont
  {Gottscholl}, \citenamefont {Diez}, \citenamefont {Soltamov}, \citenamefont
  {Kasper}, \citenamefont {Krauße}, \citenamefont {Sperlich}, \citenamefont
  {Kianinia}, \citenamefont {Bradac}, \citenamefont {Aharonovich},\ and\
  \citenamefont {Dyakonov}}]{gottscholl_spin_2021}%
  \BibitemOpen
  \bibfield  {author} {\bibinfo {author} {\bibfnamefont {A.}~\bibnamefont
  {Gottscholl}}, \bibinfo {author} {\bibfnamefont {M.}~\bibnamefont {Diez}},
  \bibinfo {author} {\bibfnamefont {V.}~\bibnamefont {Soltamov}}, \bibinfo
  {author} {\bibfnamefont {C.}~\bibnamefont {Kasper}}, \bibinfo {author}
  {\bibfnamefont {D.}~\bibnamefont {Krauße}}, \bibinfo {author} {\bibfnamefont
  {A.}~\bibnamefont {Sperlich}}, \bibinfo {author} {\bibfnamefont
  {M.}~\bibnamefont {Kianinia}}, \bibinfo {author} {\bibfnamefont
  {C.}~\bibnamefont {Bradac}}, \bibinfo {author} {\bibfnamefont
  {I.}~\bibnamefont {Aharonovich}},\ and\ \bibinfo {author} {\bibfnamefont
  {V.}~\bibnamefont {Dyakonov}},\ }\bibfield  {title} {\bibinfo {title} {Spin
  defects in {hBN} as promising temperature, pressure and magnetic field
  quantum sensors},\ }\href {https://doi.org/10.1038/s41467-021-24725-1}
  {\bibfield  {journal} {\bibinfo  {journal} {Nature Communications}\ }\textbf
  {\bibinfo {volume} {12}},\ \bibinfo {pages} {4480} (\bibinfo {year}
  {2021})}\BibitemShut {NoStop}%
\bibitem [{\citenamefont {Dolde}\ \emph {et~al.}(2011)\citenamefont {Dolde},
  \citenamefont {Fedder}, \citenamefont {Doherty}, \citenamefont {Nöbauer},
  \citenamefont {Rempp}, \citenamefont {Balasubramanian}, \citenamefont {Wolf},
  \citenamefont {Reinhard}, \citenamefont {Hollenberg}, \citenamefont
  {Jelezko},\ and\ \citenamefont {Wrachtrup}}]{dolde_electric-field_2011}%
  \BibitemOpen
  \bibfield  {author} {\bibinfo {author} {\bibfnamefont {F.}~\bibnamefont
  {Dolde}}, \bibinfo {author} {\bibfnamefont {H.}~\bibnamefont {Fedder}},
  \bibinfo {author} {\bibfnamefont {M.~W.}\ \bibnamefont {Doherty}}, \bibinfo
  {author} {\bibfnamefont {T.}~\bibnamefont {Nöbauer}}, \bibinfo {author}
  {\bibfnamefont {F.}~\bibnamefont {Rempp}}, \bibinfo {author} {\bibfnamefont
  {G.}~\bibnamefont {Balasubramanian}}, \bibinfo {author} {\bibfnamefont
  {T.}~\bibnamefont {Wolf}}, \bibinfo {author} {\bibfnamefont {F.}~\bibnamefont
  {Reinhard}}, \bibinfo {author} {\bibfnamefont {L.~C.~L.}\ \bibnamefont
  {Hollenberg}}, \bibinfo {author} {\bibfnamefont {F.}~\bibnamefont
  {Jelezko}},\ and\ \bibinfo {author} {\bibfnamefont {J.}~\bibnamefont
  {Wrachtrup}},\ }\bibfield  {title} {\bibinfo {title} {Electric-field sensing
  using single diamond spins},\ }\href {https://doi.org/10.1038/nphys1969}
  {\bibfield  {journal} {\bibinfo  {journal} {Nature Physics}\ }\textbf
  {\bibinfo {volume} {7}},\ \bibinfo {pages} {459} (\bibinfo {year}
  {2011})}\BibitemShut {NoStop}%
\bibitem [{\citenamefont {Acosta}\ \emph {et~al.}(2010)\citenamefont {Acosta},
  \citenamefont {Bauch}, \citenamefont {Ledbetter}, \citenamefont {Waxman},
  \citenamefont {Bouchard},\ and\ \citenamefont
  {Budker}}]{acosta_temperature_2010}%
  \BibitemOpen
  \bibfield  {author} {\bibinfo {author} {\bibfnamefont {V.~M.}\ \bibnamefont
  {Acosta}}, \bibinfo {author} {\bibfnamefont {E.}~\bibnamefont {Bauch}},
  \bibinfo {author} {\bibfnamefont {M.~P.}\ \bibnamefont {Ledbetter}}, \bibinfo
  {author} {\bibfnamefont {A.}~\bibnamefont {Waxman}}, \bibinfo {author}
  {\bibfnamefont {L.-S.}\ \bibnamefont {Bouchard}},\ and\ \bibinfo {author}
  {\bibfnamefont {D.}~\bibnamefont {Budker}},\ }\bibfield  {title} {\bibinfo
  {title} {Temperature {Dependence} of the {Nitrogen}-{Vacancy} {Magnetic}
  {Resonance} in {Diamond}},\ }\href
  {https://doi.org/10.1103/PhysRevLett.104.070801} {\bibfield  {journal}
  {\bibinfo  {journal} {Physical Review Letters}\ }\textbf {\bibinfo {volume}
  {104}},\ \bibinfo {pages} {070801} (\bibinfo {year} {2010})}\BibitemShut
  {NoStop}%
\bibitem [{\citenamefont {Toyli}\ \emph {et~al.}(2012)\citenamefont {Toyli},
  \citenamefont {Christle}, \citenamefont {Alkauskas}, \citenamefont {Buckley},
  \citenamefont {Van~de Walle},\ and\ \citenamefont {Awschalom}}]{Toyli_2012}%
  \BibitemOpen
  \bibfield  {author} {\bibinfo {author} {\bibfnamefont {D.~M.}\ \bibnamefont
  {Toyli}}, \bibinfo {author} {\bibfnamefont {D.~J.}\ \bibnamefont {Christle}},
  \bibinfo {author} {\bibfnamefont {A.}~\bibnamefont {Alkauskas}}, \bibinfo
  {author} {\bibfnamefont {B.~B.}\ \bibnamefont {Buckley}}, \bibinfo {author}
  {\bibfnamefont {C.~G.}\ \bibnamefont {Van~de Walle}},\ and\ \bibinfo {author}
  {\bibfnamefont {D.~D.}\ \bibnamefont {Awschalom}},\ }\bibfield  {title}
  {\bibinfo {title} {Measurement and control of single nitrogen-vacancy center
  spins above 600 k},\ }\href {https://doi.org/10.1103/PhysRevX.2.031001}
  {\bibfield  {journal} {\bibinfo  {journal} {Physical Review X}\ }\textbf
  {\bibinfo {volume} {2}},\ \bibinfo {pages} {031001} (\bibinfo {year}
  {2012})}\BibitemShut {NoStop}%
\bibitem [{\citenamefont {Doherty}\ \emph {et~al.}(2013)\citenamefont
  {Doherty}, \citenamefont {Manson}, \citenamefont {Delaney}, \citenamefont
  {Jelezko}, \citenamefont {Wrachtrup},\ and\ \citenamefont
  {Hollenberg}}]{doherty_nitrogen-vacancy_2013}%
  \BibitemOpen
  \bibfield  {author} {\bibinfo {author} {\bibfnamefont {M.~W.}\ \bibnamefont
  {Doherty}}, \bibinfo {author} {\bibfnamefont {N.~B.}\ \bibnamefont {Manson}},
  \bibinfo {author} {\bibfnamefont {P.}~\bibnamefont {Delaney}}, \bibinfo
  {author} {\bibfnamefont {F.}~\bibnamefont {Jelezko}}, \bibinfo {author}
  {\bibfnamefont {J.}~\bibnamefont {Wrachtrup}},\ and\ \bibinfo {author}
  {\bibfnamefont {L.~C.}\ \bibnamefont {Hollenberg}},\ }\bibfield  {title}
  {\bibinfo {title} {The nitrogen-vacancy colour centre in diamond},\ }\href
  {https://doi.org/10.1016/j.physrep.2013.02.001} {\bibfield  {journal}
  {\bibinfo  {journal} {Physics Reports}\ }\textbf {\bibinfo {volume} {528}},\
  \bibinfo {pages} {1} (\bibinfo {year} {2013})}\BibitemShut {NoStop}%
\bibitem [{\citenamefont {Jelezko}\ and\ \citenamefont
  {Wrachtrup}(2006)}]{jelezko_single_2006}%
  \BibitemOpen
  \bibfield  {author} {\bibinfo {author} {\bibfnamefont {F.}~\bibnamefont
  {Jelezko}}\ and\ \bibinfo {author} {\bibfnamefont {J.}~\bibnamefont
  {Wrachtrup}},\ }\bibfield  {title} {\bibinfo {title} {Single defect centres
  in diamond: {A} review},\ }\href {https://doi.org/10.1002/pssa.200671403}
  {\bibfield  {journal} {\bibinfo  {journal} {physica status solidi (a)}\
  }\textbf {\bibinfo {volume} {203}},\ \bibinfo {pages} {3207} (\bibinfo {year}
  {2006})}\BibitemShut {NoStop}%
\bibitem [{\citenamefont {Widmann}\ \emph {et~al.}(2015)\citenamefont
  {Widmann}, \citenamefont {Lee}, \citenamefont {Rendler}, \citenamefont {Son},
  \citenamefont {Fedder}, \citenamefont {Paik}, \citenamefont {Yang},
  \citenamefont {Zhao}, \citenamefont {Yang}, \citenamefont {Booker},
  \citenamefont {Denisenko}, \citenamefont {Jamali}, \citenamefont
  {Momenzadeh}, \citenamefont {Gerhardt}, \citenamefont {Ohshima},
  \citenamefont {Gali}, \citenamefont {Janzén},\ and\ \citenamefont
  {Wrachtrup}}]{widmann_coherent_2015}%
  \BibitemOpen
  \bibfield  {author} {\bibinfo {author} {\bibfnamefont {M.}~\bibnamefont
  {Widmann}}, \bibinfo {author} {\bibfnamefont {S.-Y.}\ \bibnamefont {Lee}},
  \bibinfo {author} {\bibfnamefont {T.}~\bibnamefont {Rendler}}, \bibinfo
  {author} {\bibfnamefont {N.~T.}\ \bibnamefont {Son}}, \bibinfo {author}
  {\bibfnamefont {H.}~\bibnamefont {Fedder}}, \bibinfo {author} {\bibfnamefont
  {S.}~\bibnamefont {Paik}}, \bibinfo {author} {\bibfnamefont {L.-P.}\
  \bibnamefont {Yang}}, \bibinfo {author} {\bibfnamefont {N.}~\bibnamefont
  {Zhao}}, \bibinfo {author} {\bibfnamefont {S.}~\bibnamefont {Yang}}, \bibinfo
  {author} {\bibfnamefont {I.}~\bibnamefont {Booker}}, \bibinfo {author}
  {\bibfnamefont {A.}~\bibnamefont {Denisenko}}, \bibinfo {author}
  {\bibfnamefont {M.}~\bibnamefont {Jamali}}, \bibinfo {author} {\bibfnamefont
  {S.~A.}\ \bibnamefont {Momenzadeh}}, \bibinfo {author} {\bibfnamefont
  {I.}~\bibnamefont {Gerhardt}}, \bibinfo {author} {\bibfnamefont
  {T.}~\bibnamefont {Ohshima}}, \bibinfo {author} {\bibfnamefont
  {A.}~\bibnamefont {Gali}}, \bibinfo {author} {\bibfnamefont {E.}~\bibnamefont
  {Janzén}},\ and\ \bibinfo {author} {\bibfnamefont {J.}~\bibnamefont
  {Wrachtrup}},\ }\bibfield  {title} {\bibinfo {title} {Coherent control of
  single spins in silicon carbide at room temperature},\ }\href
  {https://doi.org/10.1038/nmat4145} {\bibfield  {journal} {\bibinfo  {journal}
  {Nature Materials}\ }\textbf {\bibinfo {volume} {14}},\ \bibinfo {pages}
  {164} (\bibinfo {year} {2015})}\BibitemShut {NoStop}%
\bibitem [{\citenamefont {Koehl}\ \emph {et~al.}(2011)\citenamefont {Koehl},
  \citenamefont {Buckley}, \citenamefont {Heremans}, \citenamefont {Calusine},\
  and\ \citenamefont {Awschalom}}]{koehl_room_2011}%
  \BibitemOpen
  \bibfield  {author} {\bibinfo {author} {\bibfnamefont {W.~F.}\ \bibnamefont
  {Koehl}}, \bibinfo {author} {\bibfnamefont {B.~B.}\ \bibnamefont {Buckley}},
  \bibinfo {author} {\bibfnamefont {F.~J.}\ \bibnamefont {Heremans}}, \bibinfo
  {author} {\bibfnamefont {G.}~\bibnamefont {Calusine}},\ and\ \bibinfo
  {author} {\bibfnamefont {D.~D.}\ \bibnamefont {Awschalom}},\ }\bibfield
  {title} {\bibinfo {title} {Room temperature coherent control of defect spin
  qubits in silicon carbide},\ }\href {https://doi.org/10.1038/nature10562}
  {\bibfield  {journal} {\bibinfo  {journal} {Nature}\ }\textbf {\bibinfo
  {volume} {479}},\ \bibinfo {pages} {84} (\bibinfo {year} {2011})}\BibitemShut
  {NoStop}%
\bibitem [{\citenamefont {Gottscholl}\ \emph {et~al.}(2020)\citenamefont
  {Gottscholl}, \citenamefont {Kianinia}, \citenamefont {Soltamov},
  \citenamefont {Orlinskii}, \citenamefont {Mamin}, \citenamefont {Bradac},
  \citenamefont {Kasper}, \citenamefont {Krambrock}, \citenamefont {Sperlich},
  \citenamefont {Toth}, \citenamefont {Aharonovich},\ and\ \citenamefont
  {Dyakonov}}]{Gottscholl_2020}%
  \BibitemOpen
  \bibfield  {author} {\bibinfo {author} {\bibfnamefont {A.}~\bibnamefont
  {Gottscholl}}, \bibinfo {author} {\bibfnamefont {M.}~\bibnamefont
  {Kianinia}}, \bibinfo {author} {\bibfnamefont {V.}~\bibnamefont {Soltamov}},
  \bibinfo {author} {\bibfnamefont {S.}~\bibnamefont {Orlinskii}}, \bibinfo
  {author} {\bibfnamefont {G.}~\bibnamefont {Mamin}}, \bibinfo {author}
  {\bibfnamefont {C.}~\bibnamefont {Bradac}}, \bibinfo {author} {\bibfnamefont
  {C.}~\bibnamefont {Kasper}}, \bibinfo {author} {\bibfnamefont
  {K.}~\bibnamefont {Krambrock}}, \bibinfo {author} {\bibfnamefont
  {A.}~\bibnamefont {Sperlich}}, \bibinfo {author} {\bibfnamefont
  {M.}~\bibnamefont {Toth}}, \bibinfo {author} {\bibfnamefont {I.}~\bibnamefont
  {Aharonovich}},\ and\ \bibinfo {author} {\bibfnamefont {V.}~\bibnamefont
  {Dyakonov}},\ }\bibfield  {title} {\bibinfo {title} {Initialization and
  read-out of intrinsic spin defects in a van der waals crystal at room
  temperature},\ }\href {https://doi.org/10.1038/s41563-020-0619-6} {\bibfield
  {journal} {\bibinfo  {journal} {Nature Materials}\ }\textbf {\bibinfo
  {volume} {19}},\ \bibinfo {pages} {540} (\bibinfo {year} {2020})}\BibitemShut
  {NoStop}%
\bibitem [{\citenamefont {Gao}\ \emph {et~al.}(2021)\citenamefont {Gao},
  \citenamefont {Jiang}, \citenamefont {Llacsahuanga~Allcca}, \citenamefont
  {Shen}, \citenamefont {Sadi}, \citenamefont {Solanki}, \citenamefont {Ju},
  \citenamefont {Xu}, \citenamefont {Upadhyaya}, \citenamefont {Chen},
  \citenamefont {Bhave},\ and\ \citenamefont {Li}}]{gao_high-contrast_2021}%
  \BibitemOpen
  \bibfield  {author} {\bibinfo {author} {\bibfnamefont {X.}~\bibnamefont
  {Gao}}, \bibinfo {author} {\bibfnamefont {B.}~\bibnamefont {Jiang}}, \bibinfo
  {author} {\bibfnamefont {A.~E.}\ \bibnamefont {Llacsahuanga~Allcca}},
  \bibinfo {author} {\bibfnamefont {K.}~\bibnamefont {Shen}}, \bibinfo {author}
  {\bibfnamefont {M.~A.}\ \bibnamefont {Sadi}}, \bibinfo {author}
  {\bibfnamefont {A.~B.}\ \bibnamefont {Solanki}}, \bibinfo {author}
  {\bibfnamefont {P.}~\bibnamefont {Ju}}, \bibinfo {author} {\bibfnamefont
  {Z.}~\bibnamefont {Xu}}, \bibinfo {author} {\bibfnamefont {P.}~\bibnamefont
  {Upadhyaya}}, \bibinfo {author} {\bibfnamefont {Y.~P.}\ \bibnamefont {Chen}},
  \bibinfo {author} {\bibfnamefont {S.~A.}\ \bibnamefont {Bhave}},\ and\
  \bibinfo {author} {\bibfnamefont {T.}~\bibnamefont {Li}},\ }\bibfield
  {title} {\bibinfo {title} {High-{Contrast} {Plasmonic}-{Enhanced} {Shallow}
  {Spin} {Defects} in {Hexagonal} {Boron} {Nitride} for {Quantum} {Sensing}},\
  }\href {https://doi.org/10.1021/acs.nanolett.1c02495} {\bibfield  {journal}
  {\bibinfo  {journal} {Nano Letters}\ }\textbf {\bibinfo {volume} {21}},\
  \bibinfo {pages} {7708} (\bibinfo {year} {2021})}\BibitemShut {NoStop}%
\bibitem [{\citenamefont {Chejanovsky}\ \emph {et~al.}(2021)\citenamefont
  {Chejanovsky}, \citenamefont {Mukherjee}, \citenamefont {Geng}, \citenamefont
  {Chen}, \citenamefont {Kim}, \citenamefont {Denisenko}, \citenamefont
  {Finkler}, \citenamefont {Taniguchi}, \citenamefont {Watanabe}, \citenamefont
  {Dasari}, \citenamefont {Auburger}, \citenamefont {Gali}, \citenamefont
  {Smet},\ and\ \citenamefont {Wrachtrup}}]{chejanovsky_single-spin_2021}%
  \BibitemOpen
  \bibfield  {author} {\bibinfo {author} {\bibfnamefont {N.}~\bibnamefont
  {Chejanovsky}}, \bibinfo {author} {\bibfnamefont {A.}~\bibnamefont
  {Mukherjee}}, \bibinfo {author} {\bibfnamefont {J.}~\bibnamefont {Geng}},
  \bibinfo {author} {\bibfnamefont {Y.-C.}\ \bibnamefont {Chen}}, \bibinfo
  {author} {\bibfnamefont {Y.}~\bibnamefont {Kim}}, \bibinfo {author}
  {\bibfnamefont {A.}~\bibnamefont {Denisenko}}, \bibinfo {author}
  {\bibfnamefont {A.}~\bibnamefont {Finkler}}, \bibinfo {author} {\bibfnamefont
  {T.}~\bibnamefont {Taniguchi}}, \bibinfo {author} {\bibfnamefont
  {K.}~\bibnamefont {Watanabe}}, \bibinfo {author} {\bibfnamefont {D.~B.~R.}\
  \bibnamefont {Dasari}}, \bibinfo {author} {\bibfnamefont {P.}~\bibnamefont
  {Auburger}}, \bibinfo {author} {\bibfnamefont {A.}~\bibnamefont {Gali}},
  \bibinfo {author} {\bibfnamefont {J.~H.}\ \bibnamefont {Smet}},\ and\
  \bibinfo {author} {\bibfnamefont {J.}~\bibnamefont {Wrachtrup}},\ }\bibfield
  {title} {\bibinfo {title} {Single-spin resonance in a van der {Waals}
  embedded paramagnetic defect},\ }\href
  {https://doi.org/10.1038/s41563-021-00979-4} {\bibfield  {journal} {\bibinfo
  {journal} {Nature Materials}\ }\textbf {\bibinfo {volume} {20}},\ \bibinfo
  {pages} {1079} (\bibinfo {year} {2021})}\BibitemShut {NoStop}%
\bibitem [{\citenamefont {Stern}\ \emph {et~al.}(2022)\citenamefont {Stern},
  \citenamefont {Gu}, \citenamefont {Jarman}, \citenamefont {Eizagirre~Barker},
  \citenamefont {Mendelson}, \citenamefont {Chugh}, \citenamefont {Schott},
  \citenamefont {Tan}, \citenamefont {Sirringhaus}, \citenamefont
  {Aharonovich},\ and\ \citenamefont {Atatüre}}]{stern_room-temperature_2022}%
  \BibitemOpen
  \bibfield  {author} {\bibinfo {author} {\bibfnamefont {H.~L.}\ \bibnamefont
  {Stern}}, \bibinfo {author} {\bibfnamefont {Q.}~\bibnamefont {Gu}}, \bibinfo
  {author} {\bibfnamefont {J.}~\bibnamefont {Jarman}}, \bibinfo {author}
  {\bibfnamefont {S.}~\bibnamefont {Eizagirre~Barker}}, \bibinfo {author}
  {\bibfnamefont {N.}~\bibnamefont {Mendelson}}, \bibinfo {author}
  {\bibfnamefont {D.}~\bibnamefont {Chugh}}, \bibinfo {author} {\bibfnamefont
  {S.}~\bibnamefont {Schott}}, \bibinfo {author} {\bibfnamefont {H.~H.}\
  \bibnamefont {Tan}}, \bibinfo {author} {\bibfnamefont {H.}~\bibnamefont
  {Sirringhaus}}, \bibinfo {author} {\bibfnamefont {I.}~\bibnamefont
  {Aharonovich}},\ and\ \bibinfo {author} {\bibfnamefont {M.}~\bibnamefont
  {Atatüre}},\ }\bibfield  {title} {\bibinfo {title} {Room-temperature
  optically detected magnetic resonance of single defects in hexagonal boron
  nitride},\ }\href {https://doi.org/10.1038/s41467-022-28169-z} {\bibfield
  {journal} {\bibinfo  {journal} {Nature Communications}\ }\textbf {\bibinfo
  {volume} {13}},\ \bibinfo {pages} {618} (\bibinfo {year} {2022})}\BibitemShut
  {NoStop}%
\bibitem [{\citenamefont {Rondin}\ \emph
  {et~al.}(2014{\natexlab{b}})\citenamefont {Rondin}, \citenamefont {Tetienne},
  \citenamefont {Hingant}, \citenamefont {Roch}, \citenamefont {Maletinsky},\
  and\ \citenamefont {Jacques}}]{rondin_magnetometry_2014}%
  \BibitemOpen
  \bibfield  {author} {\bibinfo {author} {\bibfnamefont {L.}~\bibnamefont
  {Rondin}}, \bibinfo {author} {\bibfnamefont {J.-P.}\ \bibnamefont
  {Tetienne}}, \bibinfo {author} {\bibfnamefont {T.}~\bibnamefont {Hingant}},
  \bibinfo {author} {\bibfnamefont {J.-F.}\ \bibnamefont {Roch}}, \bibinfo
  {author} {\bibfnamefont {P.}~\bibnamefont {Maletinsky}},\ and\ \bibinfo
  {author} {\bibfnamefont {V.}~\bibnamefont {Jacques}},\ }\bibfield  {title}
  {\bibinfo {title} {Magnetometry with nitrogen-vacancy defects in diamond},\
  }\href {https://doi.org/10.1088/0034-4885/77/5/056503} {\bibfield  {journal}
  {\bibinfo  {journal} {Reports on Progress in Physics}\ }\textbf {\bibinfo
  {volume} {77}},\ \bibinfo {pages} {056503} (\bibinfo {year}
  {2014}{\natexlab{b}})}\BibitemShut {NoStop}%
\bibitem [{\citenamefont {Mathur}\ \emph {et~al.}(2022)\citenamefont {Mathur},
  \citenamefont {Mukherjee}, \citenamefont {Gao}, \citenamefont {Luo},
  \citenamefont {McCullian}, \citenamefont {Li}, \citenamefont {Vamivakas},\
  and\ \citenamefont {Fuchs}}]{mathur_excited-state_2022}%
  \BibitemOpen
  \bibfield  {author} {\bibinfo {author} {\bibfnamefont {N.}~\bibnamefont
  {Mathur}}, \bibinfo {author} {\bibfnamefont {A.}~\bibnamefont {Mukherjee}},
  \bibinfo {author} {\bibfnamefont {X.}~\bibnamefont {Gao}}, \bibinfo {author}
  {\bibfnamefont {J.}~\bibnamefont {Luo}}, \bibinfo {author} {\bibfnamefont
  {B.~A.}\ \bibnamefont {McCullian}}, \bibinfo {author} {\bibfnamefont
  {T.}~\bibnamefont {Li}}, \bibinfo {author} {\bibfnamefont {A.~N.}\
  \bibnamefont {Vamivakas}},\ and\ \bibinfo {author} {\bibfnamefont {G.~D.}\
  \bibnamefont {Fuchs}},\ }\bibfield  {title} {\bibinfo {title} {Excited-state
  spin-resonance spectroscopy of {$V_B^-$} defect centers in hexagonal boron
  nitride},\ }\href {https://doi.org/10.1038/s41467-022-30772-z} {\bibfield
  {journal} {\bibinfo  {journal} {Nature Communications}\ }\textbf {\bibinfo
  {volume} {13}},\ \bibinfo {pages} {3233} (\bibinfo {year}
  {2022})}\BibitemShut {NoStop}%
\bibitem [{\citenamefont {Mu}\ \emph {et~al.}(2022)\citenamefont {Mu},
  \citenamefont {Cai}, \citenamefont {Chen}, \citenamefont {Kenny},
  \citenamefont {Jiang}, \citenamefont {Ru}, \citenamefont {Lyu}, \citenamefont
  {Koh}, \citenamefont {Liu}, \citenamefont {Aharonovich},\ and\ \citenamefont
  {Gao}}]{mu_excited-state_2022}%
  \BibitemOpen
  \bibfield  {author} {\bibinfo {author} {\bibfnamefont {Z.}~\bibnamefont
  {Mu}}, \bibinfo {author} {\bibfnamefont {H.}~\bibnamefont {Cai}}, \bibinfo
  {author} {\bibfnamefont {D.}~\bibnamefont {Chen}}, \bibinfo {author}
  {\bibfnamefont {J.}~\bibnamefont {Kenny}}, \bibinfo {author} {\bibfnamefont
  {Z.}~\bibnamefont {Jiang}}, \bibinfo {author} {\bibfnamefont
  {S.}~\bibnamefont {Ru}}, \bibinfo {author} {\bibfnamefont {X.}~\bibnamefont
  {Lyu}}, \bibinfo {author} {\bibfnamefont {T.~S.}\ \bibnamefont {Koh}},
  \bibinfo {author} {\bibfnamefont {X.}~\bibnamefont {Liu}}, \bibinfo {author}
  {\bibfnamefont {I.}~\bibnamefont {Aharonovich}},\ and\ \bibinfo {author}
  {\bibfnamefont {W.}~\bibnamefont {Gao}},\ }\bibfield  {title} {\bibinfo
  {title} {Excited-{State} {Optically} {Detected} {Magnetic} {Resonance} of
  {Spin} {Defects} in {Hexagonal} {Boron} {Nitride}},\ }\href
  {https://doi.org/10.1103/PhysRevLett.128.216402} {\bibfield  {journal}
  {\bibinfo  {journal} {Physical Review Letters}\ }\textbf {\bibinfo {volume}
  {128}},\ \bibinfo {pages} {216402} (\bibinfo {year} {2022})}\BibitemShut
  {NoStop}%
\bibitem [{\citenamefont {Reimers}\ \emph {et~al.}(2020)\citenamefont
  {Reimers}, \citenamefont {Shen}, \citenamefont {Kianinia}, \citenamefont
  {Bradac}, \citenamefont {Aharonovich}, \citenamefont {Ford},\ and\
  \citenamefont {Piecuch}}]{reimers_photoluminescence_2020}%
  \BibitemOpen
  \bibfield  {author} {\bibinfo {author} {\bibfnamefont {J.~R.}\ \bibnamefont
  {Reimers}}, \bibinfo {author} {\bibfnamefont {J.}~\bibnamefont {Shen}},
  \bibinfo {author} {\bibfnamefont {M.}~\bibnamefont {Kianinia}}, \bibinfo
  {author} {\bibfnamefont {C.}~\bibnamefont {Bradac}}, \bibinfo {author}
  {\bibfnamefont {I.}~\bibnamefont {Aharonovich}}, \bibinfo {author}
  {\bibfnamefont {M.~J.}\ \bibnamefont {Ford}},\ and\ \bibinfo {author}
  {\bibfnamefont {P.}~\bibnamefont {Piecuch}},\ }\bibfield  {title} {\bibinfo
  {title} {Photoluminescence, photophysics, and photochemistry of the
  {$V_B^{-}$} defect in hexagonal boron nitride},\ }\href
  {https://doi.org/10.1103/PhysRevB.102.144105} {\bibfield  {journal} {\bibinfo
   {journal} {Physical Review B}\ }\textbf {\bibinfo {volume} {102}},\ \bibinfo
  {pages} {144105} (\bibinfo {year} {2020})}\BibitemShut {NoStop}%
\bibitem [{\citenamefont {Xu}\ \emph {et~al.}(2023)\citenamefont {Xu},
  \citenamefont {Solanki}, \citenamefont {Sychev}, \citenamefont {Gao},
  \citenamefont {Peana}, \citenamefont {Baburin}, \citenamefont {Pagadala},
  \citenamefont {Martin}, \citenamefont {Chowdhury}, \citenamefont {Chen},
  \citenamefont {Taniguchi}, \citenamefont {Watanabe}, \citenamefont
  {Rodionov}, \citenamefont {Kildishev}, \citenamefont {Li}, \citenamefont
  {Upadhyaya}, \citenamefont {Boltasseva},\ and\ \citenamefont
  {Shalaev}}]{xu_greatly_2023}%
  \BibitemOpen
  \bibfield  {author} {\bibinfo {author} {\bibfnamefont {X.}~\bibnamefont
  {Xu}}, \bibinfo {author} {\bibfnamefont {A.~B.}\ \bibnamefont {Solanki}},
  \bibinfo {author} {\bibfnamefont {D.}~\bibnamefont {Sychev}}, \bibinfo
  {author} {\bibfnamefont {X.}~\bibnamefont {Gao}}, \bibinfo {author}
  {\bibfnamefont {S.}~\bibnamefont {Peana}}, \bibinfo {author} {\bibfnamefont
  {A.~S.}\ \bibnamefont {Baburin}}, \bibinfo {author} {\bibfnamefont
  {K.}~\bibnamefont {Pagadala}}, \bibinfo {author} {\bibfnamefont {Z.~O.}\
  \bibnamefont {Martin}}, \bibinfo {author} {\bibfnamefont {S.~N.}\
  \bibnamefont {Chowdhury}}, \bibinfo {author} {\bibfnamefont {Y.~P.}\
  \bibnamefont {Chen}}, \bibinfo {author} {\bibfnamefont {T.}~\bibnamefont
  {Taniguchi}}, \bibinfo {author} {\bibfnamefont {K.}~\bibnamefont {Watanabe}},
  \bibinfo {author} {\bibfnamefont {I.~A.}\ \bibnamefont {Rodionov}}, \bibinfo
  {author} {\bibfnamefont {A.~V.}\ \bibnamefont {Kildishev}}, \bibinfo {author}
  {\bibfnamefont {T.}~\bibnamefont {Li}}, \bibinfo {author} {\bibfnamefont
  {P.}~\bibnamefont {Upadhyaya}}, \bibinfo {author} {\bibfnamefont
  {A.}~\bibnamefont {Boltasseva}},\ and\ \bibinfo {author} {\bibfnamefont
  {V.~M.}\ \bibnamefont {Shalaev}},\ }\bibfield  {title} {\bibinfo {title}
  {Greatly {Enhanced} {Emission} from {Spin} {Defects} in {Hexagonal} {Boron}
  {Nitride} {Enabled} by a {Low}-{Loss} {Plasmonic} {Nanocavity}},\ }\href
  {https://doi.org/10.1021/acs.nanolett.2c03100} {\bibfield  {journal}
  {\bibinfo  {journal} {Nano Letters}\ }\textbf {\bibinfo {volume} {23}},\
  \bibinfo {pages} {25} (\bibinfo {year} {2023})}\BibitemShut {NoStop}%
\bibitem [{\citenamefont {Qian}\ \emph {et~al.}(2022)\citenamefont {Qian},
  \citenamefont {Villafañe}, \citenamefont {Schalk}, \citenamefont {Astakhov},
  \citenamefont {Kentsch}, \citenamefont {Helm}, \citenamefont {Soubelet},
  \citenamefont {Wilson}, \citenamefont {Rizzato}, \citenamefont {Mohr},
  \citenamefont {Holleitner}, \citenamefont {Bucher}, \citenamefont {Stier},\
  and\ \citenamefont {Finley}}]{qian_unveiling_2022}%
  \BibitemOpen
  \bibfield  {author} {\bibinfo {author} {\bibfnamefont {C.}~\bibnamefont
  {Qian}}, \bibinfo {author} {\bibfnamefont {V.}~\bibnamefont {Villafañe}},
  \bibinfo {author} {\bibfnamefont {M.}~\bibnamefont {Schalk}}, \bibinfo
  {author} {\bibfnamefont {G.~V.}\ \bibnamefont {Astakhov}}, \bibinfo {author}
  {\bibfnamefont {U.}~\bibnamefont {Kentsch}}, \bibinfo {author} {\bibfnamefont
  {M.}~\bibnamefont {Helm}}, \bibinfo {author} {\bibfnamefont {P.}~\bibnamefont
  {Soubelet}}, \bibinfo {author} {\bibfnamefont {N.~P.}\ \bibnamefont
  {Wilson}}, \bibinfo {author} {\bibfnamefont {R.}~\bibnamefont {Rizzato}},
  \bibinfo {author} {\bibfnamefont {S.}~\bibnamefont {Mohr}}, \bibinfo {author}
  {\bibfnamefont {A.~W.}\ \bibnamefont {Holleitner}}, \bibinfo {author}
  {\bibfnamefont {D.~B.}\ \bibnamefont {Bucher}}, \bibinfo {author}
  {\bibfnamefont {A.~V.}\ \bibnamefont {Stier}},\ and\ \bibinfo {author}
  {\bibfnamefont {J.~J.}\ \bibnamefont {Finley}},\ }\bibfield  {title}
  {\bibinfo {title} {Unveiling the {Zero}-{Phonon} {Line} of the {Boron}
  {Vacancy} {Center} by {Cavity}-{Enhanced} {Emission}},\ }\href
  {https://doi.org/10.1021/acs.nanolett.2c00739} {\bibfield  {journal}
  {\bibinfo  {journal} {Nano Letters}\ }\textbf {\bibinfo {volume} {22}},\
  \bibinfo {pages} {5137} (\bibinfo {year} {2022})}\BibitemShut {NoStop}%
\bibitem [{\citenamefont {Lukin}\ \emph {et~al.}(2020)\citenamefont {Lukin},
  \citenamefont {Dory}, \citenamefont {Guidry}, \citenamefont {Yang},
  \citenamefont {Mishra}, \citenamefont {Trivedi}, \citenamefont {Radulaski},
  \citenamefont {Sun}, \citenamefont {Vercruysse}, \citenamefont {Ahn},\ and\
  \citenamefont {Vučković}}]{lukin_4h-silicon-carbide--insulator_2020}%
  \BibitemOpen
  \bibfield  {author} {\bibinfo {author} {\bibfnamefont {D.~M.}\ \bibnamefont
  {Lukin}}, \bibinfo {author} {\bibfnamefont {C.}~\bibnamefont {Dory}},
  \bibinfo {author} {\bibfnamefont {M.~A.}\ \bibnamefont {Guidry}}, \bibinfo
  {author} {\bibfnamefont {K.~Y.}\ \bibnamefont {Yang}}, \bibinfo {author}
  {\bibfnamefont {S.~D.}\ \bibnamefont {Mishra}}, \bibinfo {author}
  {\bibfnamefont {R.}~\bibnamefont {Trivedi}}, \bibinfo {author} {\bibfnamefont
  {M.}~\bibnamefont {Radulaski}}, \bibinfo {author} {\bibfnamefont
  {S.}~\bibnamefont {Sun}}, \bibinfo {author} {\bibfnamefont {D.}~\bibnamefont
  {Vercruysse}}, \bibinfo {author} {\bibfnamefont {G.~H.}\ \bibnamefont
  {Ahn}},\ and\ \bibinfo {author} {\bibfnamefont {J.}~\bibnamefont
  {Vučković}},\ }\bibfield  {title} {\bibinfo {title}
  {{4H}-silicon-carbide-on-insulator for integrated quantum and nonlinear
  photonics},\ }\href {https://doi.org/10.1038/s41566-019-0556-6} {\bibfield
  {journal} {\bibinfo  {journal} {Nature Photonics}\ }\textbf {\bibinfo
  {volume} {14}},\ \bibinfo {pages} {330} (\bibinfo {year} {2020})}\BibitemShut
  {NoStop}%
\bibitem [{\citenamefont {Wang}\ \emph {et~al.}(2017)\citenamefont {Wang},
  \citenamefont {Zhou}, \citenamefont {Zhang}, \citenamefont {Liu},
  \citenamefont {Li}, \citenamefont {Li}, \citenamefont {Liu}, \citenamefont
  {Wang},\ and\ \citenamefont {Gao}}]{wang_efficient_2017}%
  \BibitemOpen
  \bibfield  {author} {\bibinfo {author} {\bibfnamefont {J.}~\bibnamefont
  {Wang}}, \bibinfo {author} {\bibfnamefont {Y.}~\bibnamefont {Zhou}}, \bibinfo
  {author} {\bibfnamefont {X.}~\bibnamefont {Zhang}}, \bibinfo {author}
  {\bibfnamefont {F.}~\bibnamefont {Liu}}, \bibinfo {author} {\bibfnamefont
  {Y.}~\bibnamefont {Li}}, \bibinfo {author} {\bibfnamefont {K.}~\bibnamefont
  {Li}}, \bibinfo {author} {\bibfnamefont {Z.}~\bibnamefont {Liu}}, \bibinfo
  {author} {\bibfnamefont {G.}~\bibnamefont {Wang}},\ and\ \bibinfo {author}
  {\bibfnamefont {W.}~\bibnamefont {Gao}},\ }\bibfield  {title} {\bibinfo
  {title} {Efficient {Generation} of an {Array} of {Single} {Silicon}-{Vacancy}
  {Defects} in {Silicon} {Carbide}},\ }\href
  {https://doi.org/10.1103/PhysRevApplied.7.064021} {\bibfield  {journal}
  {\bibinfo  {journal} {Physical Review Applied}\ }\textbf {\bibinfo {volume}
  {7}},\ \bibinfo {pages} {064021} (\bibinfo {year} {2017})}\BibitemShut
  {NoStop}%
\bibitem [{\citenamefont {Wang}\ \emph {et~al.}(2020)\citenamefont {Wang},
  \citenamefont {Yan}, \citenamefont {Li}, \citenamefont {Liu}, \citenamefont
  {Liu}, \citenamefont {Guo}, \citenamefont {Guo}, \citenamefont {Zhou},
  \citenamefont {Cui}, \citenamefont {Wang}, \citenamefont {Zhou},
  \citenamefont {Xu}, \citenamefont {Xu}, \citenamefont {Li},\ and\
  \citenamefont {Guo}}]{wang_coherent_2020}%
  \BibitemOpen
  \bibfield  {author} {\bibinfo {author} {\bibfnamefont {J.-F.}\ \bibnamefont
  {Wang}}, \bibinfo {author} {\bibfnamefont {F.-F.}\ \bibnamefont {Yan}},
  \bibinfo {author} {\bibfnamefont {Q.}~\bibnamefont {Li}}, \bibinfo {author}
  {\bibfnamefont {Z.-H.}\ \bibnamefont {Liu}}, \bibinfo {author} {\bibfnamefont
  {H.}~\bibnamefont {Liu}}, \bibinfo {author} {\bibfnamefont {G.-P.}\
  \bibnamefont {Guo}}, \bibinfo {author} {\bibfnamefont {L.-P.}\ \bibnamefont
  {Guo}}, \bibinfo {author} {\bibfnamefont {X.}~\bibnamefont {Zhou}}, \bibinfo
  {author} {\bibfnamefont {J.-M.}\ \bibnamefont {Cui}}, \bibinfo {author}
  {\bibfnamefont {J.}~\bibnamefont {Wang}}, \bibinfo {author} {\bibfnamefont
  {Z.-Q.}\ \bibnamefont {Zhou}}, \bibinfo {author} {\bibfnamefont {X.-Y.}\
  \bibnamefont {Xu}}, \bibinfo {author} {\bibfnamefont {J.-S.}\ \bibnamefont
  {Xu}}, \bibinfo {author} {\bibfnamefont {C.-F.}\ \bibnamefont {Li}},\ and\
  \bibinfo {author} {\bibfnamefont {G.-C.}\ \bibnamefont {Guo}},\ }\bibfield
  {title} {\bibinfo {title} {Coherent {Control} of {Nitrogen}-{Vacancy}
  {Center} {Spins} in {Silicon} {Carbide} at {Room} {Temperature}},\ }\href
  {https://doi.org/10.1103/PhysRevLett.124.223601} {\bibfield  {journal}
  {\bibinfo  {journal} {Physical Review Letters}\ }\textbf {\bibinfo {volume}
  {124}},\ \bibinfo {pages} {223601} (\bibinfo {year} {2020})}\BibitemShut
  {NoStop}%
\bibitem [{\citenamefont {Burk}\ \emph {et~al.}(1999)\citenamefont {Burk},
  \citenamefont {O'Loughlin}, \citenamefont {Siergiej}, \citenamefont
  {Agarwal}, \citenamefont {Sriram}, \citenamefont {Clarke}, \citenamefont
  {MacMillan}, \citenamefont {Balakrishna},\ and\ \citenamefont
  {Brandt}}]{burk_sic_1999}%
  \BibitemOpen
  \bibfield  {author} {\bibinfo {author} {\bibfnamefont {A.~A.}\ \bibnamefont
  {Burk}}, \bibinfo {author} {\bibfnamefont {M.~J.}\ \bibnamefont
  {O'Loughlin}}, \bibinfo {author} {\bibfnamefont {R.~R.}\ \bibnamefont
  {Siergiej}}, \bibinfo {author} {\bibfnamefont {A.~K.}\ \bibnamefont
  {Agarwal}}, \bibinfo {author} {\bibfnamefont {S.}~\bibnamefont {Sriram}},
  \bibinfo {author} {\bibfnamefont {R.~C.}\ \bibnamefont {Clarke}}, \bibinfo
  {author} {\bibfnamefont {M.~F.}\ \bibnamefont {MacMillan}}, \bibinfo {author}
  {\bibfnamefont {V.}~\bibnamefont {Balakrishna}},\ and\ \bibinfo {author}
  {\bibfnamefont {C.~D.}\ \bibnamefont {Brandt}},\ }\bibfield  {title}
  {\bibinfo {title} {{SiC} and {GaN} wide bandgap semiconductor materials and
  devices},\ }\href {https://doi.org/10.1016/S0038-1101(99)00089-1} {\bibfield
  {journal} {\bibinfo  {journal} {Solid-State Electronics}\ }\textbf {\bibinfo
  {volume} {43}},\ \bibinfo {pages} {1459} (\bibinfo {year}
  {1999})}\BibitemShut {NoStop}%
\bibitem [{\citenamefont {Milligan}\ \emph {et~al.}(2007)\citenamefont
  {Milligan}, \citenamefont {Sheppard}, \citenamefont {Pribble}, \citenamefont
  {Wu}, \citenamefont {Muller},\ and\ \citenamefont
  {Palmour}}]{milligan_sic_2007}%
  \BibitemOpen
  \bibfield  {author} {\bibinfo {author} {\bibfnamefont {J.~W.}\ \bibnamefont
  {Milligan}}, \bibinfo {author} {\bibfnamefont {S.}~\bibnamefont {Sheppard}},
  \bibinfo {author} {\bibfnamefont {W.}~\bibnamefont {Pribble}}, \bibinfo
  {author} {\bibfnamefont {Y.-F.}\ \bibnamefont {Wu}}, \bibinfo {author}
  {\bibfnamefont {G.}~\bibnamefont {Muller}},\ and\ \bibinfo {author}
  {\bibfnamefont {J.~W.}\ \bibnamefont {Palmour}},\ }\bibfield  {title}
  {\bibinfo {title} {{SiC} and {GaN} {Wide} {Bandgap} {Device} {Technology}
  {Overview}},\ }in\ \href {https://doi.org/10.1109/RADAR.2007.374395} {\emph
  {\bibinfo {booktitle} {2007 {IEEE} {Radar} {Conference}}}}\ (\bibinfo {year}
  {2007})\ pp.\ \bibinfo {pages} {960--964}\BibitemShut {NoStop}%
\bibitem [{\citenamefont {Chen}\ \emph {et~al.}(2017)\citenamefont {Chen},
  \citenamefont {Häberlen}, \citenamefont {Lidow}, \citenamefont {Tsai},
  \citenamefont {Ueda}, \citenamefont {Uemoto},\ and\ \citenamefont
  {Wu}}]{chen_gan--si_2017}%
  \BibitemOpen
  \bibfield  {author} {\bibinfo {author} {\bibfnamefont {K.~J.}\ \bibnamefont
  {Chen}}, \bibinfo {author} {\bibfnamefont {O.}~\bibnamefont {Häberlen}},
  \bibinfo {author} {\bibfnamefont {A.}~\bibnamefont {Lidow}}, \bibinfo
  {author} {\bibfnamefont {C.~l.}\ \bibnamefont {Tsai}}, \bibinfo {author}
  {\bibfnamefont {T.}~\bibnamefont {Ueda}}, \bibinfo {author} {\bibfnamefont
  {Y.}~\bibnamefont {Uemoto}},\ and\ \bibinfo {author} {\bibfnamefont
  {Y.}~\bibnamefont {Wu}},\ }\bibfield  {title} {\bibinfo {title}
  {{GaN}-on-{Si} {Power} {Technology}: {Devices} and {Applications}},\ }\href
  {https://doi.org/10.1109/TED.2017.2657579} {\bibfield  {journal} {\bibinfo
  {journal} {IEEE Transactions on Electron Devices}\ }\textbf {\bibinfo
  {volume} {64}},\ \bibinfo {pages} {779} (\bibinfo {year} {2017})}\BibitemShut
  {NoStop}%
\bibitem [{\citenamefont {Mishra}\ \emph {et~al.}(2002)\citenamefont {Mishra},
  \citenamefont {Parikh},\ and\ \citenamefont {Wu}}]{mishra_algangan_2002}%
  \BibitemOpen
  \bibfield  {author} {\bibinfo {author} {\bibfnamefont {U.}~\bibnamefont
  {Mishra}}, \bibinfo {author} {\bibfnamefont {P.}~\bibnamefont {Parikh}},\
  and\ \bibinfo {author} {\bibfnamefont {Y.-F.}\ \bibnamefont {Wu}},\
  }\bibfield  {title} {\bibinfo {title} {{AlGaN}/{GaN} {HEMTs}-an overview of
  device operation and applications},\ }\href
  {https://doi.org/10.1109/JPROC.2002.1021567} {\bibfield  {journal} {\bibinfo
  {journal} {Proceedings of the IEEE}\ }\textbf {\bibinfo {volume} {90}},\
  \bibinfo {pages} {1022} (\bibinfo {year} {2002})}\BibitemShut {NoStop}%
\bibitem [{\citenamefont {Berhane}\ \emph {et~al.}(2017)\citenamefont
  {Berhane}, \citenamefont {Jeong}, \citenamefont {Bodrog}, \citenamefont
  {Fiedler}, \citenamefont {Schröder}, \citenamefont {Triviño}, \citenamefont
  {Palacios}, \citenamefont {Gali}, \citenamefont {Toth}, \citenamefont
  {Englund},\ and\ \citenamefont {Aharonovich}}]{berhane_bright_2017}%
  \BibitemOpen
  \bibfield  {author} {\bibinfo {author} {\bibfnamefont {A.~M.}\ \bibnamefont
  {Berhane}}, \bibinfo {author} {\bibfnamefont {K.-Y.}\ \bibnamefont {Jeong}},
  \bibinfo {author} {\bibfnamefont {Z.}~\bibnamefont {Bodrog}}, \bibinfo
  {author} {\bibfnamefont {S.}~\bibnamefont {Fiedler}}, \bibinfo {author}
  {\bibfnamefont {T.}~\bibnamefont {Schröder}}, \bibinfo {author}
  {\bibfnamefont {N.~V.}\ \bibnamefont {Triviño}}, \bibinfo {author}
  {\bibfnamefont {T.}~\bibnamefont {Palacios}}, \bibinfo {author}
  {\bibfnamefont {A.}~\bibnamefont {Gali}}, \bibinfo {author} {\bibfnamefont
  {M.}~\bibnamefont {Toth}}, \bibinfo {author} {\bibfnamefont {D.}~\bibnamefont
  {Englund}},\ and\ \bibinfo {author} {\bibfnamefont {I.}~\bibnamefont
  {Aharonovich}},\ }\bibfield  {title} {\bibinfo {title} {Bright
  {Room}-{Temperature} {Single}-{Photon} {Emission} from {Defects} in {Gallium}
  {Nitride}},\ }\href {https://doi.org/10.1002/adma.201605092} {\bibfield
  {journal} {\bibinfo  {journal} {Advanced Materials}\ }\textbf {\bibinfo
  {volume} {29}},\ \bibinfo {pages} {1605092} (\bibinfo {year}
  {2017})}\BibitemShut {NoStop}%
\bibitem [{\citenamefont {Berhane}\ \emph {et~al.}(2018)\citenamefont
  {Berhane}, \citenamefont {Jeong}, \citenamefont {Bradac}, \citenamefont
  {Walsh}, \citenamefont {Englund}, \citenamefont {Toth},\ and\ \citenamefont
  {Aharonovich}}]{berhane_photophysics_2018}%
  \BibitemOpen
  \bibfield  {author} {\bibinfo {author} {\bibfnamefont {A.~M.}\ \bibnamefont
  {Berhane}}, \bibinfo {author} {\bibfnamefont {K.-Y.}\ \bibnamefont {Jeong}},
  \bibinfo {author} {\bibfnamefont {C.}~\bibnamefont {Bradac}}, \bibinfo
  {author} {\bibfnamefont {M.}~\bibnamefont {Walsh}}, \bibinfo {author}
  {\bibfnamefont {D.}~\bibnamefont {Englund}}, \bibinfo {author} {\bibfnamefont
  {M.}~\bibnamefont {Toth}},\ and\ \bibinfo {author} {\bibfnamefont
  {I.}~\bibnamefont {Aharonovich}},\ }\bibfield  {title} {\bibinfo {title}
  {Photophysics of {GaN} single-photon emitters in the visible spectral
  range},\ }\href {https://doi.org/10.1103/PhysRevB.97.165202} {\bibfield
  {journal} {\bibinfo  {journal} {Physical Review B}\ }\textbf {\bibinfo
  {volume} {97}},\ \bibinfo {pages} {165202} (\bibinfo {year}
  {2018})}\BibitemShut {NoStop}%
\bibitem [{\citenamefont {Geng}\ \emph {et~al.}(2023)\citenamefont {Geng},
  \citenamefont {Luo}, \citenamefont {Van~Deurzen}, \citenamefont {Xing},
  \citenamefont {Jena}, \citenamefont {Fuchs},\ and\ \citenamefont
  {Rana}}]{geng_dephasing_2023}%
  \BibitemOpen
  \bibfield  {author} {\bibinfo {author} {\bibfnamefont {Y.}~\bibnamefont
  {Geng}}, \bibinfo {author} {\bibfnamefont {J.}~\bibnamefont {Luo}}, \bibinfo
  {author} {\bibfnamefont {L.}~\bibnamefont {Van~Deurzen}}, \bibinfo {author}
  {\bibfnamefont {H.}~\bibnamefont {Xing}}, \bibinfo {author} {\bibfnamefont
  {D.}~\bibnamefont {Jena}}, \bibinfo {author} {\bibfnamefont {G.~D.}\
  \bibnamefont {Fuchs}},\ and\ \bibinfo {author} {\bibfnamefont
  {F.}~\bibnamefont {Rana}},\ }\bibfield  {title} {\bibinfo {title} {Dephasing
  by optical phonons in {GaN} defect single-photon emitters},\ }\href
  {https://doi.org/10.1038/s41598-023-35003-z} {\bibfield  {journal} {\bibinfo
  {journal} {Scientific Reports}\ }\textbf {\bibinfo {volume} {13}},\ \bibinfo
  {pages} {8678} (\bibinfo {year} {2023})}\BibitemShut {NoStop}%
\bibitem [{\citenamefont {Epstein}\ \emph {et~al.}(2005)\citenamefont
  {Epstein}, \citenamefont {Mendoza}, \citenamefont {Kato},\ and\ \citenamefont
  {Awschalom}}]{epstein_anisotropic_2005}%
  \BibitemOpen
  \bibfield  {author} {\bibinfo {author} {\bibfnamefont {R.~J.}\ \bibnamefont
  {Epstein}}, \bibinfo {author} {\bibfnamefont {F.~M.}\ \bibnamefont
  {Mendoza}}, \bibinfo {author} {\bibfnamefont {Y.~K.}\ \bibnamefont {Kato}},\
  and\ \bibinfo {author} {\bibfnamefont {D.~D.}\ \bibnamefont {Awschalom}},\
  }\bibfield  {title} {\bibinfo {title} {Anisotropic interactions of a single
  spin and dark-spin spectroscopy in diamond},\ }\href
  {https://doi.org/10.1038/nphys141} {\bibfield  {journal} {\bibinfo  {journal}
  {Nature Physics}\ }\textbf {\bibinfo {volume} {1}},\ \bibinfo {pages} {94}
  (\bibinfo {year} {2005})}\BibitemShut {NoStop}%
\bibitem [{\citenamefont {Exarhos}\ \emph {et~al.}(2019)\citenamefont
  {Exarhos}, \citenamefont {Hopper}, \citenamefont {Patel}, \citenamefont
  {Doherty},\ and\ \citenamefont
  {Bassett}}]{exarhos_magnetic-field-dependent_2019}%
  \BibitemOpen
  \bibfield  {author} {\bibinfo {author} {\bibfnamefont {A.~L.}\ \bibnamefont
  {Exarhos}}, \bibinfo {author} {\bibfnamefont {D.~A.}\ \bibnamefont {Hopper}},
  \bibinfo {author} {\bibfnamefont {R.~N.}\ \bibnamefont {Patel}}, \bibinfo
  {author} {\bibfnamefont {M.~W.}\ \bibnamefont {Doherty}},\ and\ \bibinfo
  {author} {\bibfnamefont {L.~C.}\ \bibnamefont {Bassett}},\ }\bibfield
  {title} {\bibinfo {title} {Magnetic-field-dependent quantum emission in
  hexagonal boron nitride at room temperature},\ }\href
  {https://doi.org/10.1038/s41467-018-08185-8} {\bibfield  {journal} {\bibinfo
  {journal} {Nature Communications}\ }\textbf {\bibinfo {volume} {10}},\
  \bibinfo {pages} {222} (\bibinfo {year} {2019})}\BibitemShut {NoStop}%
\bibitem [{\citenamefont {Lee}\ \emph {et~al.}(2013)\citenamefont {Lee},
  \citenamefont {Widmann}, \citenamefont {Rendler}, \citenamefont {Doherty},
  \citenamefont {Babinec}, \citenamefont {Yang}, \citenamefont {Eyer},
  \citenamefont {Siyushev}, \citenamefont {Hausmann}, \citenamefont {Loncar},
  \citenamefont {Bodrog}, \citenamefont {Gali}, \citenamefont {Manson},
  \citenamefont {Fedder},\ and\ \citenamefont {Wrachtrup}}]{lee_readout_2013}%
  \BibitemOpen
  \bibfield  {author} {\bibinfo {author} {\bibfnamefont {S.-Y.}\ \bibnamefont
  {Lee}}, \bibinfo {author} {\bibfnamefont {M.}~\bibnamefont {Widmann}},
  \bibinfo {author} {\bibfnamefont {T.}~\bibnamefont {Rendler}}, \bibinfo
  {author} {\bibfnamefont {M.~W.}\ \bibnamefont {Doherty}}, \bibinfo {author}
  {\bibfnamefont {T.~M.}\ \bibnamefont {Babinec}}, \bibinfo {author}
  {\bibfnamefont {S.}~\bibnamefont {Yang}}, \bibinfo {author} {\bibfnamefont
  {M.}~\bibnamefont {Eyer}}, \bibinfo {author} {\bibfnamefont {P.}~\bibnamefont
  {Siyushev}}, \bibinfo {author} {\bibfnamefont {B.~J.~M.}\ \bibnamefont
  {Hausmann}}, \bibinfo {author} {\bibfnamefont {M.}~\bibnamefont {Loncar}},
  \bibinfo {author} {\bibfnamefont {Z.}~\bibnamefont {Bodrog}}, \bibinfo
  {author} {\bibfnamefont {A.}~\bibnamefont {Gali}}, \bibinfo {author}
  {\bibfnamefont {N.~B.}\ \bibnamefont {Manson}}, \bibinfo {author}
  {\bibfnamefont {H.}~\bibnamefont {Fedder}},\ and\ \bibinfo {author}
  {\bibfnamefont {J.}~\bibnamefont {Wrachtrup}},\ }\bibfield  {title} {\bibinfo
  {title} {Readout and control of a single nuclear spin with a metastable
  electron spin ancilla},\ }\href {https://doi.org/10.1038/nnano.2013.104}
  {\bibfield  {journal} {\bibinfo  {journal} {Nature Nanotechnology}\ }\textbf
  {\bibinfo {volume} {8}},\ \bibinfo {pages} {487} (\bibinfo {year}
  {2013})}\BibitemShut {NoStop}%
\end{thebibliography}%

\pagebreak
\widetext
\setcounter{equation}{0}
\setcounter{figure}{0}
\setcounter{table}{0}
\setcounter{page}{1}
\makeatletter
\renewcommand{\theequation}{S\arabic{equation}}
\renewcommand{\thefigure}{S\arabic{figure}}
\renewcommand{\bibnumfmt}[1]{[S#1]}
\renewcommand{\citenumfont}[1]{S#1}
\begin{center}
\textbf{\large Supplemental Materials: \\Room temperature optically detected magnetic resonance of single spins in GaN}
\end{center}
\section{Measurement Setup}
\begin{figure}[!h]
    \centering
    \includegraphics[width=0.6\linewidth]{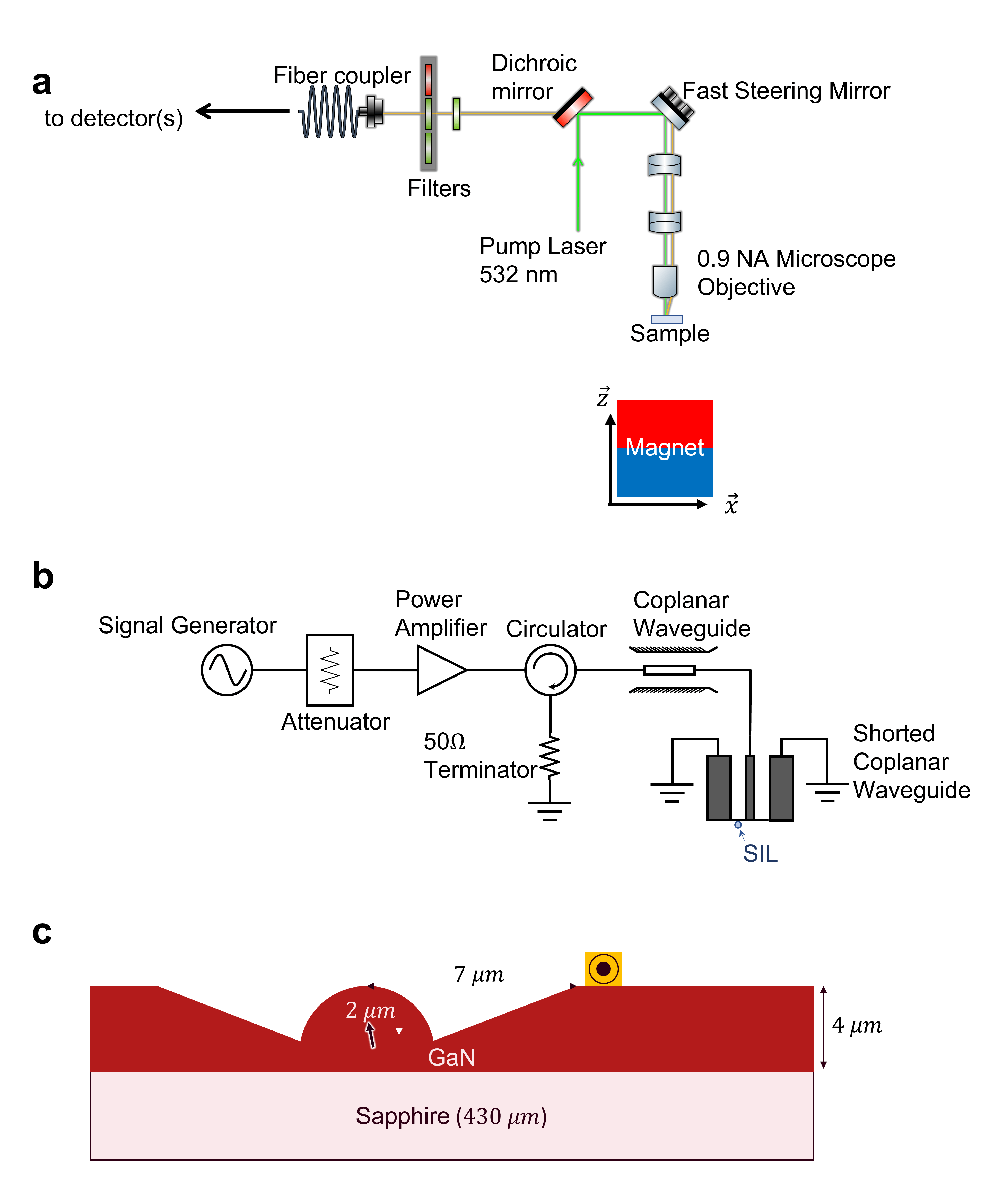}
    \caption{
    \textbf{Setup schematics.} (a) The scanning laser confocal microscope setup used in all measurements. (b) Microwave signal chain for driving spin resonances. (c) Cross-sectional schematic of the microwire placement with respect to the SIL.
    }
    \label{fig:extFig:setup_schematics}
\end{figure}

Figure~\ref{fig:extFig:setup_schematics}(a) shows our home-built scanning laser confocal microscope setup and Fig.~\ref{fig:extFig:setup_schematics}(b) the microwave signal chain for driving the spin resonance in both continuous-wave-ODMR (cw-ODMR) and pulsed-ODMR. 

We use a focused ion beam to carve out a solid-immersion-lens (SIL) around defects of interest. Using these structures, we are able to excite and collect the PL efficiently with modest laser power, around 15--20~$\mu$W. 

\begin{figure}[!h]
    \centering
    \includegraphics[width=0.5\linewidth]{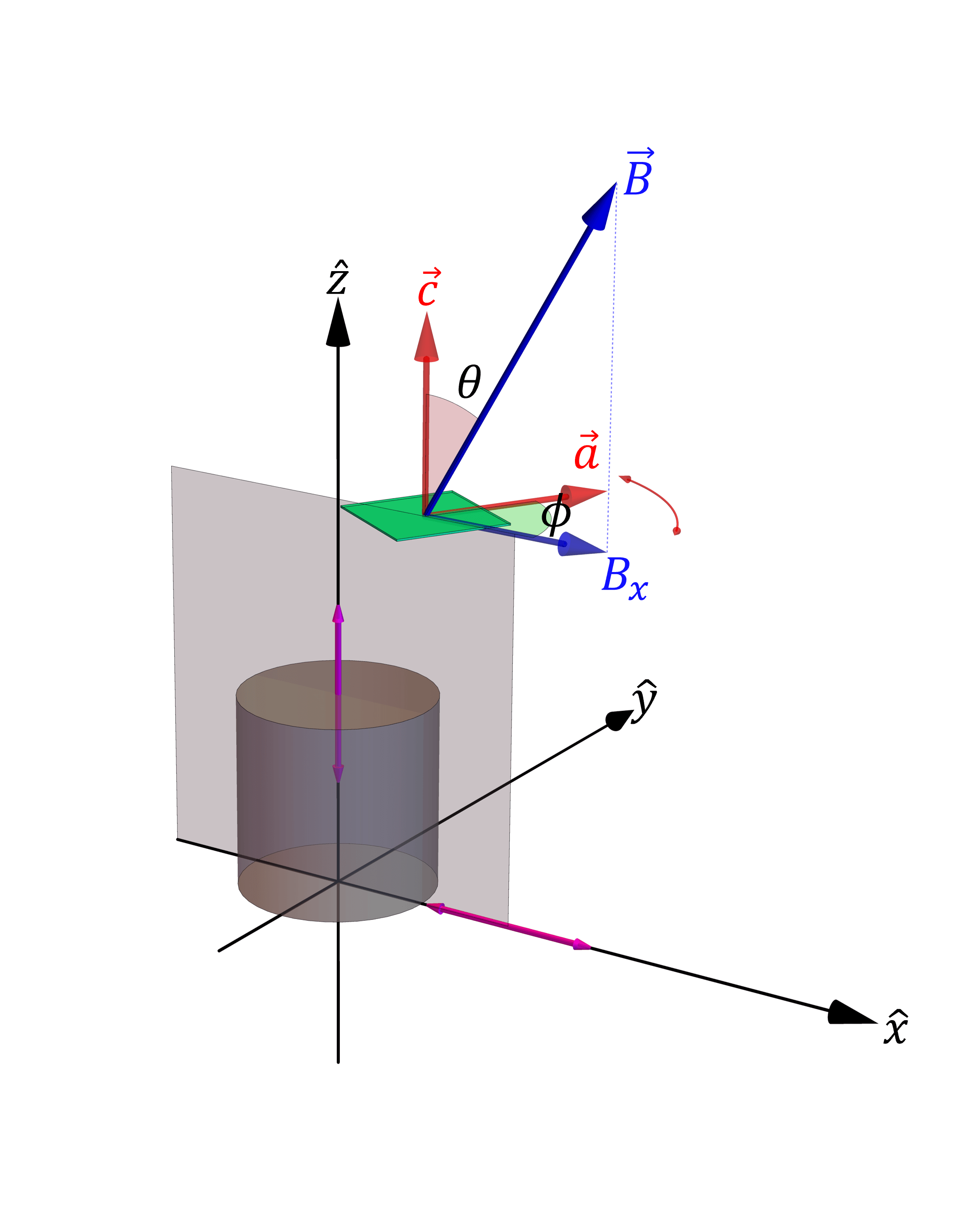}
    \caption{
    \textbf{Magnet setup scheme.}
    The magnet $z$-axis, the GaN $c$-axis, and the optical axis are parallel to each other. The optical axis coincides with the $c$-axis of the sample at the defects.
    We move the magnet in both the $x$ and $z$-axis with motorized translation stages to control the polar angle $\theta$ and we rotate the sample about the $c$-axis to change the azimuth $\phi$, the angle between the magnetic field onto the sample plane and the crystal lattice vector $\vec{a}$. The straight purple double arrows represent the directions of translations of the magnet, and the curved purple double arrow represents the rotational degree of freedom of the sample.
    }
    \label{fig:extFig:magnetGeometry}
\end{figure}

We apply magnetic fields to the sample using a 50.4-cm-long 50.4-cm-diameter permanent neodymium iron boron magnet mounted on a two-axis translation stage. 
To precisely control the magnetic field angle at the defect location, we first align the symmetry axis of the cylindrical magnet to the optical axis of the setup and mount the GaN sample so that its $c$-axis coincides with the optical axis.
We now can tilt the magnetic field with respect to the sample $c$-axis by translating the magnet in $x$-axis as shown in Fig.~\ref{fig:extFig:magnetGeometry}. This angle is labeled as the polar angle $\theta$. When measuring the $\theta$ dependence of ODMR contrast, we also translate the magnet in $z$-direction such that the magnetic field norm is maintained within $100$~G.

We rotate the sample to change the magnetic field projected onto the $c$-plane and we label the angle from this projection to the lattice crystal $a$-axis as the azimuth $\phi$. 
We calibrate the magnetic field as a function of the magnet position at the sample position with a Hall probe. The resulting measured field distribution matches closely with a calculation of the magnetic field from a cylindrical magnet of these dimensions.

To apply microwave magnetic excitation, we lithographically print a shorted coplanar waveguide on the surface of GaN.  
The short consists of a copper wire with a 1~$\mu$m-side square cross-section (see Fig.~\ref{fig:extFig:setup_schematics}c). The wire is patterned about 7~$\mu$m away from the center of the SILs to avoid the SIL structure.

\newpage

\section{Optical properties of defects studied}
\begin{figure}[!hbpt]
    \centering
    \includegraphics[width=1\linewidth]{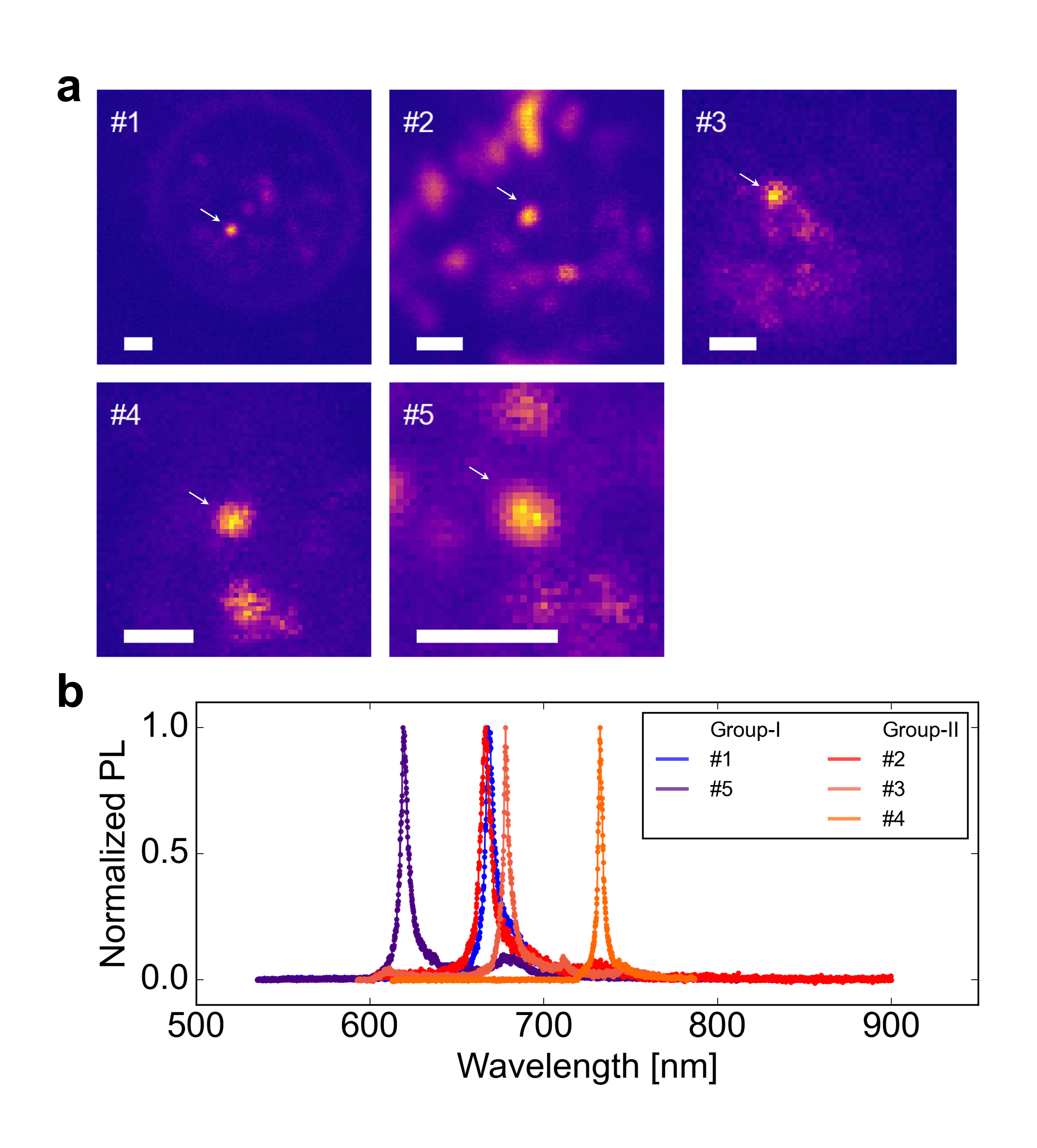}
    \caption{
    \textbf{Optical properties of defects studied.}
    (a) The PL map and (b) the PL spectra of defects measured in this work. The scale bars in (a) are 1~$\mu$m long.
    }
    \label{fig:extFig:allPLMapSpectra}
\end{figure}

Figure~\ref{fig:extFig:allPLMapSpectra} shows the PL images and the corresponding PL spectra of the defects studied in this work. The linewidths range from 3~nm to 10~nm at room temperature with most of the photons emission in the zero-phonon-line. 

\begin{figure}[!hpbt]
    \centering
    \includegraphics[width=0.8\linewidth]{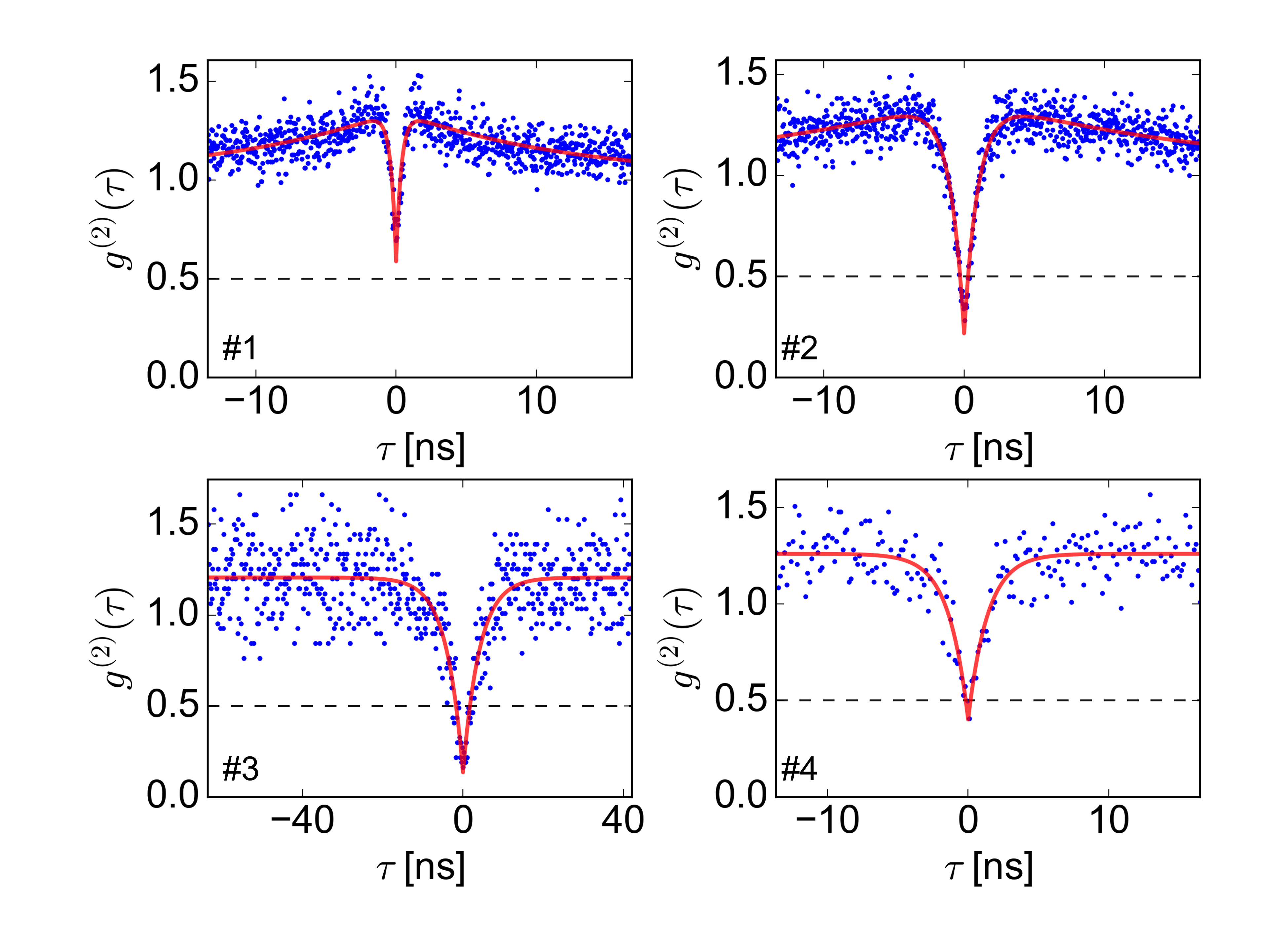}
    \caption{
    \textbf{Photon auto-correlation ($g^{(2)}$) of defects studied.}
     Plots of $g^{(2)}(\tau)$ for defects \#1 -- \#4).
    }
    \label{fig:extFig:allG2}
\end{figure}

Figure~\ref{fig:extFig:allG2} shows the photon auto-correlation $g^{(2)}$ measurements of defects \#1--4.  We expect all of the defects that we investigated are single defects. Defects \#2--4 all display $g^{(2)}(0) < 0.5$, which is strong evidence that they are in fact single photon emitters. The $g^{(2)}(0)$ of defect \#1 does not dip below 0.5. However, we note that the decay constant associated with the central dip of $g^{(2)}$ is 350~ps, which means that in the presence of the APD time jitter, also about 350~ps, $g^{(2)}(0)$ is limited by the instrument response, and thus this measurement is consistent with our assignment of defect \#1 as a single emitter. 
Defect \#5 is absent beyond the magneto-PL measurement, because it is no longer optically active through photo-bleaching or some other mechanism.


\section{Angle Dependent ODMR}
\begin{figure}[!hbpt]
    \centering
    \includegraphics[width=1\linewidth]{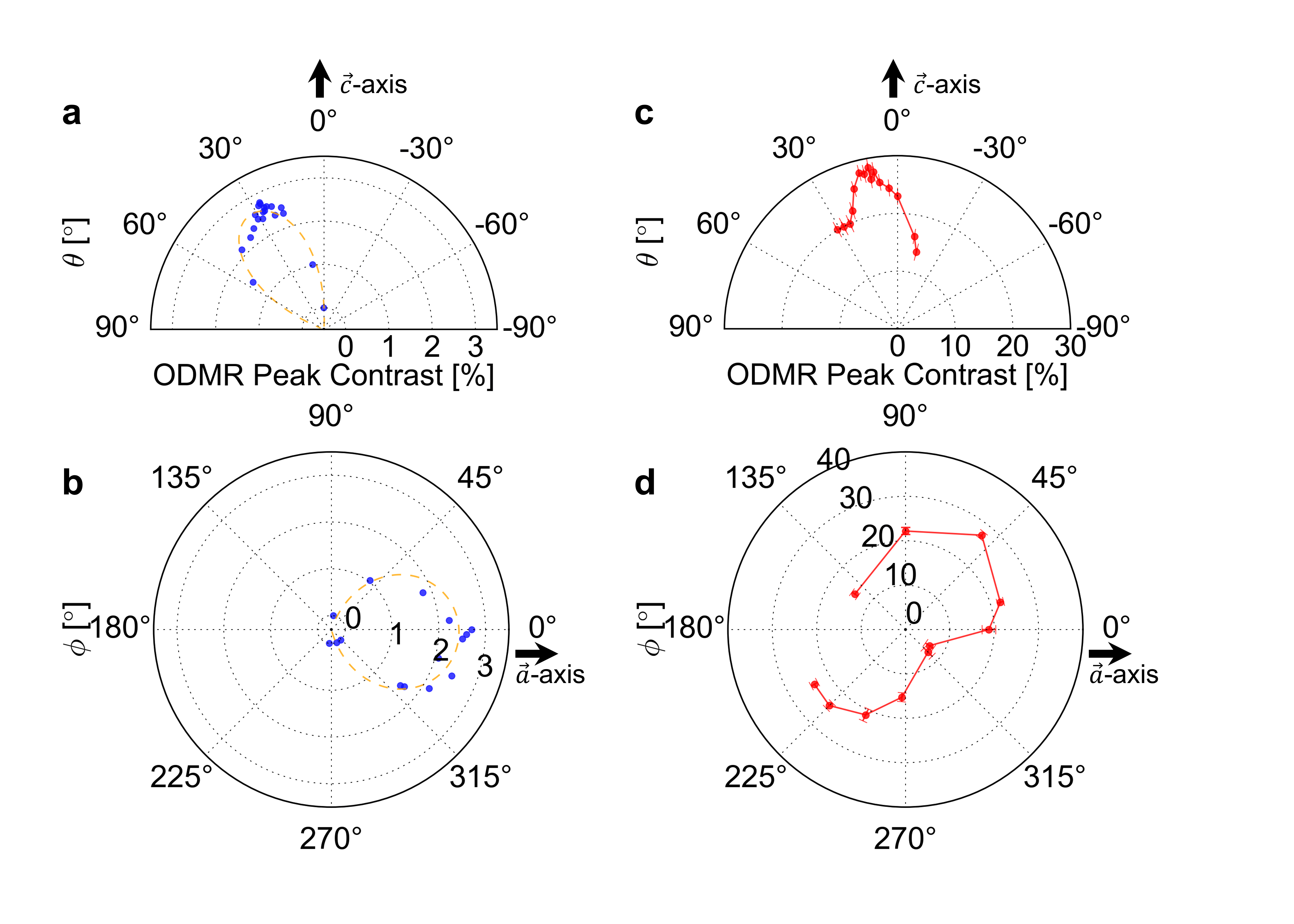}
    \caption{
    \textbf{Angle-dependent ODMR contrast of defects \#1 and \#2.}
    ODMR contrasts of defects (a) \#1  and (c) \#2 are measured when the magnetic field is tilted with respect to the $c$-axis of GaN crystal. (b) and (d) show the ODMR contrasts as a function of the in-plane component of the magnetic field. 
    The measurements are done when we keep $\theta=27^\circ$ and $\theta=72^\circ$ for defects \#1 and \#2, respectively. 
    The dashed lines in (a) and (b) are guides to the eye.
    }
    \label{fig:extFig:defect1and2AngleData}
\end{figure}

Figure~\ref{fig:extFig:defect1and2AngleData}(a) and (b) show the angle dependencies of defect \#1. 
The ODMR signals are only above measurement sensitivity when $\theta$ and $\phi$ are within a small window. 

For defect \#2, the dependencies are weaker. We find an ODMR signal maximum when the magnetic field points at $\theta=10^\circ$ from the $c$-axis  and when its projection forms a $\phi=60^\circ$ with an $a$-axis of the crystal as seen in Fig.~\ref{fig:extFig:defect1and2AngleData}(c) and (d), respectively. 

The optimal directions for spin quantization axes are visualized in the main text. Neither connects a lattice site to its nearest few neighbors.

\section{cw-ODMR as a function of magnetic field}
\begin{figure}[!hpbt]
    \centering
    \includegraphics[width=1\linewidth]{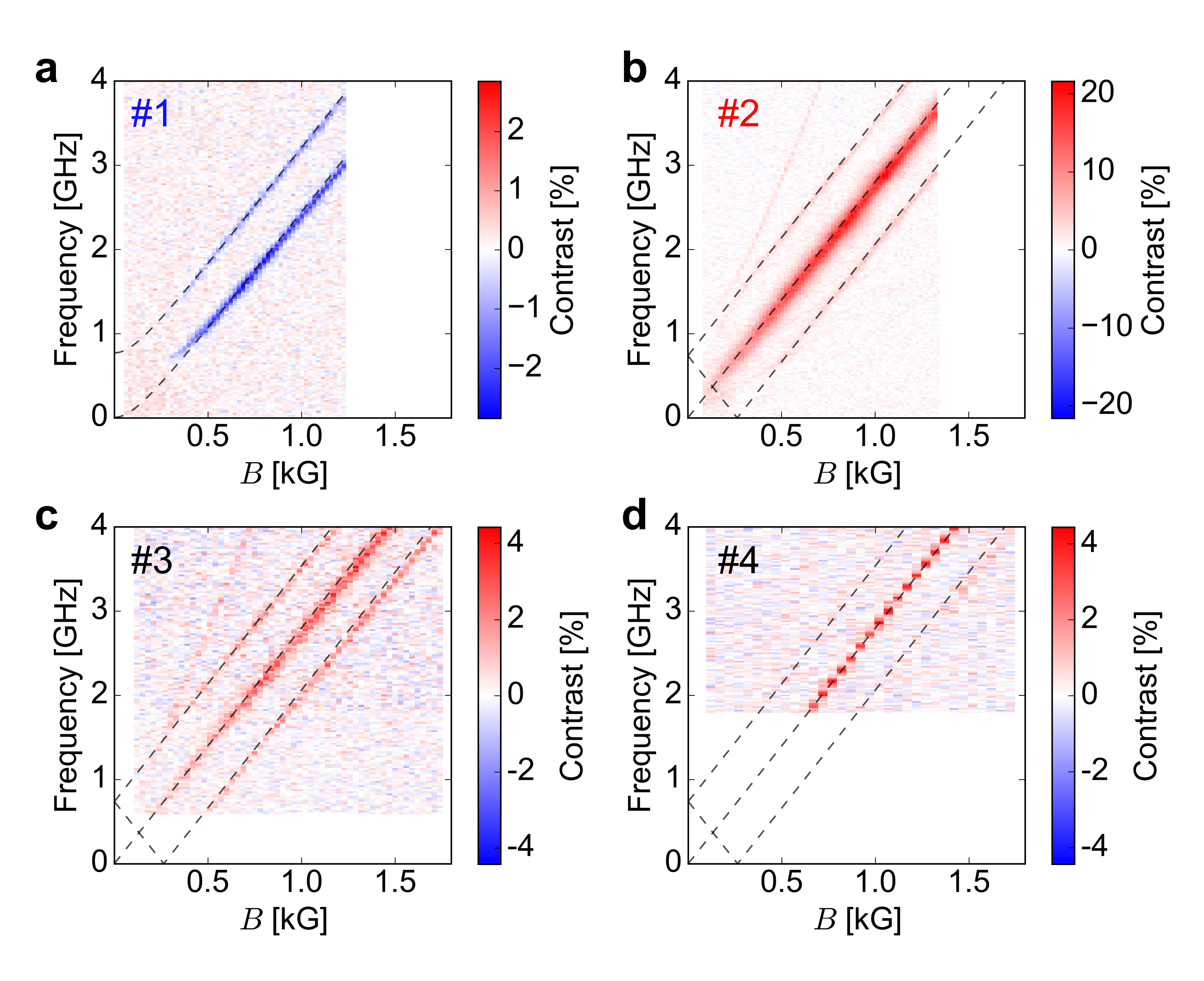}
    \caption{
    \textbf{Magnetic field dependent ODMR of four defects.} We align the magnetic field to produce the largest contrast resonance and record the ODMR signals as a function of magnetic fields. Note that defects (\#2--4 or b--d) in group-II essentially have the same spin structures. The fit lines in (a) are transition frequencies from a spin-1 model with $D\approx E\approx 389$~MHz. The fit lines in (b--d) are transition frequencies fit to three spin resonances that disperse with $g=2$ from the data of defect \#2, with $\Delta m_s=1$ from a spin-3/2 model with $E=0$ and $D=368$~MHz. 
    }
    \label{fig:extFig:allODMRSpectra}
\end{figure}

Figure~\ref{fig:extFig:allODMRSpectra} shows cw-ODMR traces as a function of the magnetic field that is aligned with the angle of highest ODMR contrast for each. For these measurements, the magnetic field is re-calibrated in this position, and then moved only along the axial direction to change the magnetic field amplitude but not its direction during the measurement. 

These data show two groups of spin resonance responses, as discussed in the main text. 
Defect \#1 shows positive contrast with only two transition frequencies. Thus we label it group-I because it shares the same magneto-PL with defect \#5. 
We fit the group-I defect \#1 data with a spin-1 model, which results in a pair of commensurate axial and transversal zero field splitting parameters, $D\approx E\approx 389$~MHz.
The minimal model deviates from the experimental data at low field ($B<0.4$~kG) where complicated dynamics and spin mixing are not captured by the Hamiltonian discussed in the main text.
Defects \#2--4 each show positive ODMR contrast, and have essentially the same spin structures within experimental uncertainties as seen from the fit lines shown in Fig.~\ref{fig:extFig:allODMRSpectra}(b--d). 
The group-II line fit is done to the defect \#2 data with a spin-3/2 model, accounting only for the spin resonances that disperse with $g=2$. The fit results in zero field  splitting parameters of $E=0$ and $D=369$~MHz. Once we obtained the fit to defect \#2 data, we simply overlay it over the data of defects \#3 and \#4.
We note that they do not all have the same contrast, which could be due to differences in their local environments.

\section{Timings for pulsed measurements}
\begin{figure}[!hpbt]
    \centering
    \includegraphics[width=0.75\linewidth]{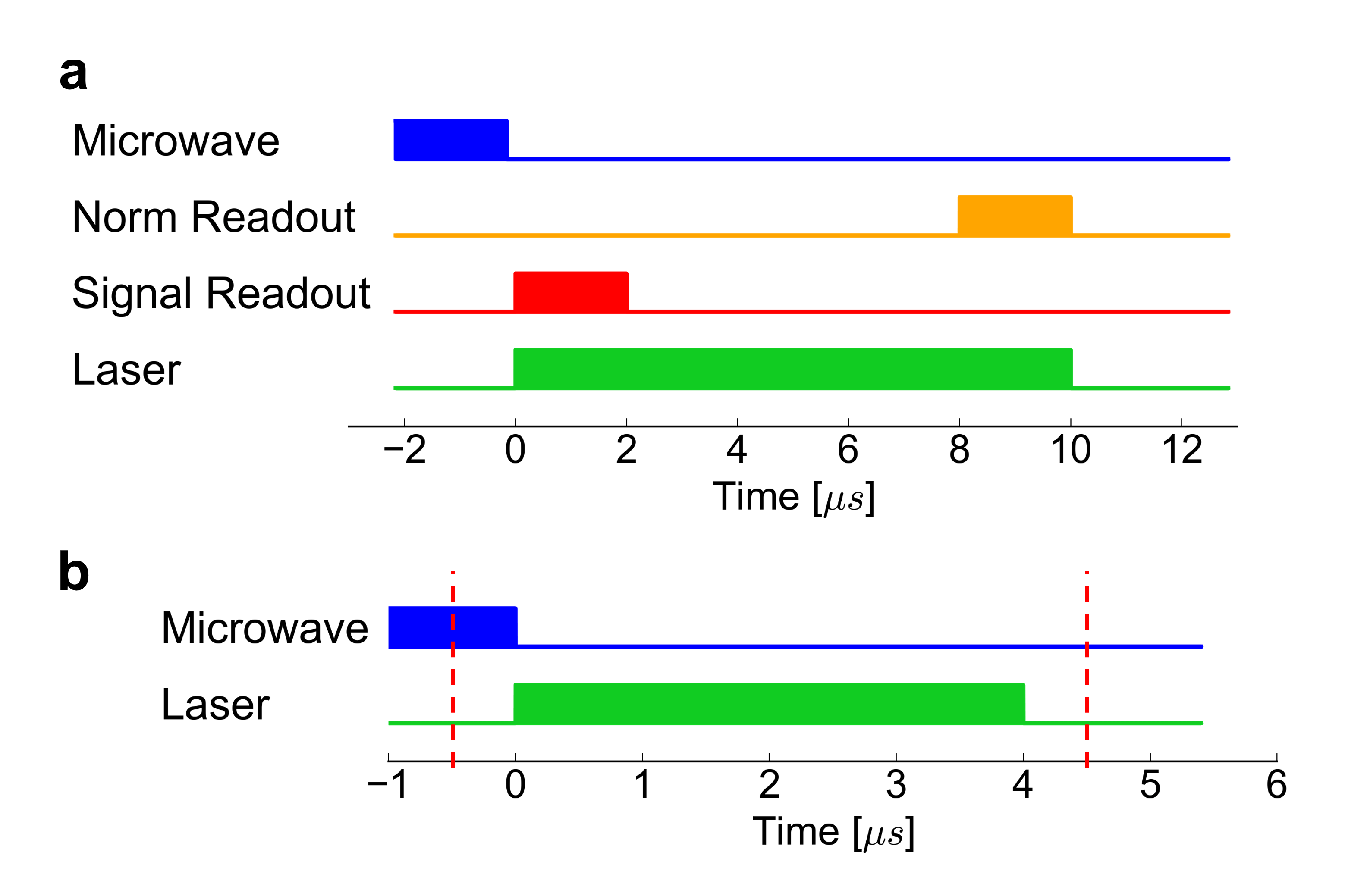}
    \caption{
    \textbf{Timing diagram for pulsed measurements.} The diagrams illustrate the timings in a cycle for (a) pulsed ODMR and (b) time-resolved PL measurements. Red dashed lines in (b) enclose the time-resolved PL data acquisition window presented in the main text.
    }
    \label{fig:extFig:pulsedTiming}
\end{figure}

Figure~\ref{fig:extFig:pulsedTiming} shows the details of the pulse timing used in pulsed ODMR and time-resolved PL measurement.

\section{Photostability of GaN defects}
\begin{figure}[h!]
    \centering
    \includegraphics[width=1\linewidth]{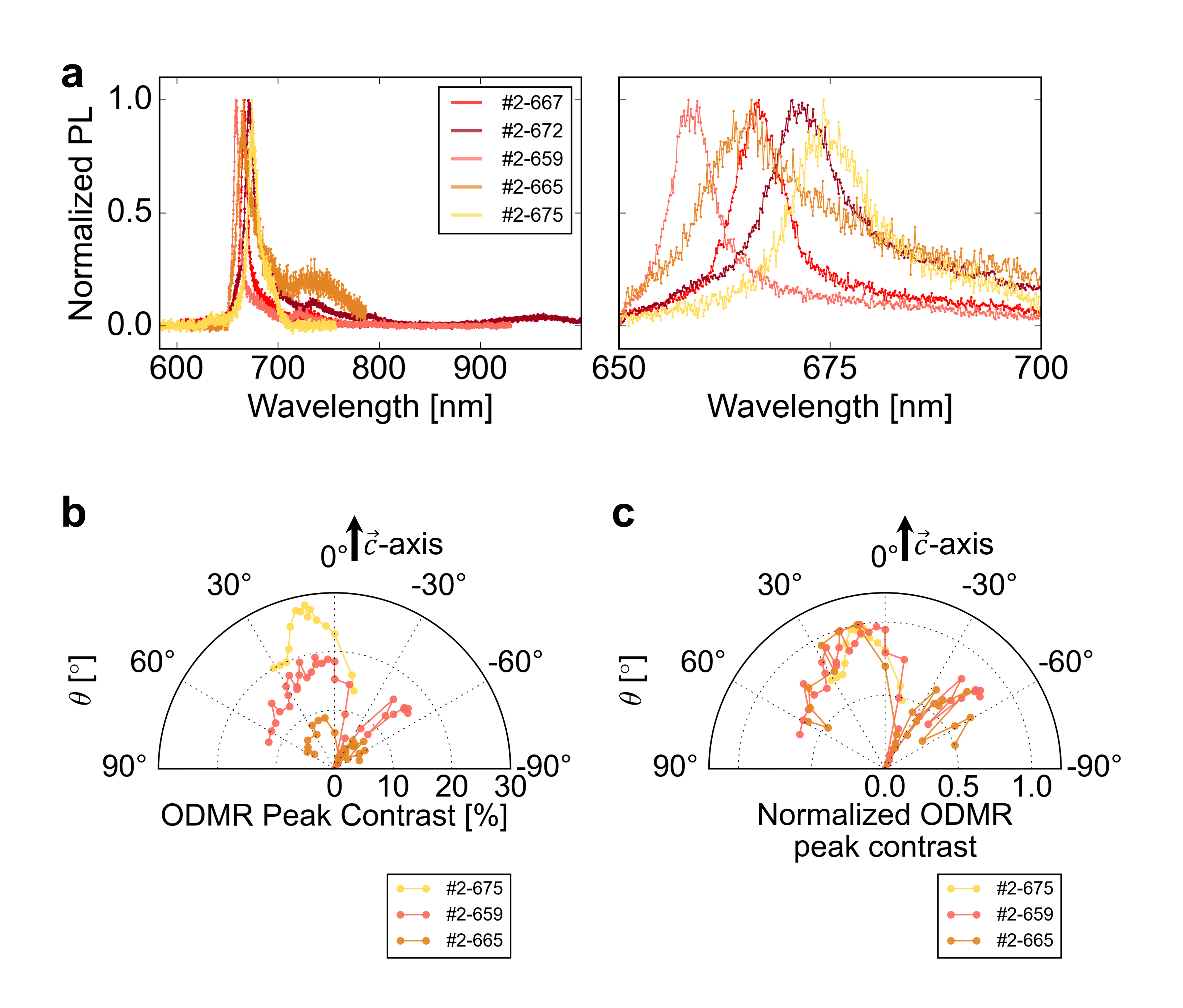}
    \caption{
    \textbf{Defect \#2 photo-instability.} (a) The observed different PL spectra of defect \#2 and zoomed-in replica near the zero-phonon-line wavelengths. We label the variants by their ZPL wavelengths.
    (b) shows the peak contrast of the largest ODMR signal as a function of magnetic field angles $\theta$.
    (c) shows the same set of data with the contrast normalized to the maximum ODMR of the same variant.
    }
    \label{fig:extFig:defectNO2PLspectraAndAngle}
\end{figure}

Some GaN defects are stable for a long time, and some behave differently.
For example, defect \#1 has shown the same PL spectra for more than a year, whereas defect \#2 has shown discrete PL spectra changes after a few weeks of study as seen in Fig.~\ref{fig:extFig:defectNO2PLspectraAndAngle}(a). 
We note defect \#2 could change into and out of a particular variant on a timescale of tens of minutes to hours.
Defect \#5 photobleached after approximately a few hours of 20~$\mu$W laser excitation.

ODMR contrast changes accompany the PL spectra changes observed in defect \#2, although the ODMR transition frequencies and Hamiltonian appear not to change. 
We are able to capture some of this behavior by monitoring the PL spectra during the ODMR measurements.
We find that the $\theta$ dependence of the ODMR signal on the magnetic field does not change significantly, as seen in Fig.~\ref{fig:extFig:defectNO2PLspectraAndAngle}(b,c).
It is possible that the laser excitation traps or removes charges in the defect's local environment, and the charge environment changes the PL spectra but not the spin structure. 

\end{document}